\documentclass[useAMS,usenatbib,letterpaper,usegraphicx]{mn2e}
\usepackage{times,amssymb,hyperref,subfigure,epsfig,aas_macros}
\usepackage[totalwidth=480pt,totalheight=680pt,layoutvoffset=0.5cm]{geometry}

\begin{document}

\title[A Circumbinary Polar Ring Debris Disk]{99 Herculis: Host to a Circumbinary
  Polar-ring Debris Disk}

\author[G. M. Kennedy et. al.]{G. M. Kennedy\thanks{Email:
    \href{mailto:gkennedy@ast.cam.ac.uk}{gkennedy@ast.cam.ac.uk}}$^1$, M. C. Wyatt$^1$,
  B. Sibthorpe$^2$, G. Duch\^ene$^{3,4}$, P. Kalas$^4$, \newauthor
  B. C. Matthews$^{5,6}$, J. S. Greaves$^7$, K. Y. L. Su$^8$, M. P. Fitzgerald$^{9,10}$ \vspace{0.2cm} \\
  $^1$ Institute of Astronomy, University of Cambridge, Madingley Road, Cambridge CB3 0HA, UK\\
  $^2$ UK Astronomy Technology Center, Royal Observatory, Blackford Hill, Edinburgh EH9
  3HJ, UK\\
  $^3$ Department of Astronomy, University of California, B-20 Hearst Field Annex,
  Berkeley, CA
  94720-3411, USA\\
  $^4$ Laboratoire d'Astrophysique, Observatoire de Grenoble, Universit\'e J. Fourier,
  CNRS, France\\
  $^5$ Herzberg Institute of Astrophysics, National Research Council Canada, 5071 West
  Saanich Road., Victoria, BC, Canada, V9E 2E7, Canada\\
  $^6$ University of Victoria, Finnerty Road, Victoria, BC, V8W 3P6, Canada\\
  $^7$ School of Physics and Astronomy, University of St Andrews, North Haugh, St Andrews, Fife KY16 9SS, UK \\
  $^8$ Steward Observatory, University of Arizona, 933 North Cherry Avenue, Tucson, AZ
  85721, USA\\
  $^9$ Institute of Geophysics and Planetary Physics, Lawrence Livermore National
  Laboratory, L-413, 7000 East Avenue, Livermore, CA 94550, USA\\
  $^{10}$ Department of Physics and Astronomy, UCLA, Los Angeles, CA 90095-1547, USA }

\maketitle

\begin{abstract}
  We present resolved \emph{Herschel} images of a circumbinary debris disk in the 99
  Herculis system. The primary is a late F-type star. The binary orbit is well
  characterised and we conclude that the disk is misaligned with the binary plane. Two
  different models can explain the observed structure. The first model is a ring of polar
  orbits that move in a plane perpendicular to the binary pericenter direction. We favour
  this interpretation because it includes the effect of secular perturbations and the
  disk can survive for Gyr timescales. The second model is a misaligned ring. Because
  there is an ambiguity in the orientation of the ring, which could be reflected in the
  sky plane, this ring either has near-polar orbits similar to the first model, or has a
  30 degree misalignment. The misaligned ring, interpreted as the result of a recent
  collision, is shown to be implausible from constraints on the collisional and dynamical
  evolution. Because disk+star systems with separations similar to 99 Herculis should
  form coplanar, possible formation scenarios involve either a close stellar encounter or
  binary exchange in the presence of circumstellar and/or circumbinary disks.  Discovery
  and characterisation of systems like 99 Herculis will help understand processes that
  result in planetary system misalignment around both single and multiple stars.
\end{abstract}


\section{Introduction}\label{s:intro}

The \emph{Herschel} Key Program DEBRIS (Dust Emission via a Bias free Reconnaissance in
the Infrared/Submillimeter) has observed a large sample of nearby stars to discover and
characterise extrasolar analogues to the Solar System's asteroid and Kuiper belts,
collectively known as ``debris disks.'' The 3.5m \emph{Herschel} mirror diameter provides
6-7'' resolution at 70-100$\mu$m \citep{2010A&A...518L...1P}, and as a consequence our
survey has resolved many disks around stars in the Solar neighbourhood for the first time
\citep{2010A&A...518L.135M,2011MNRAS.417.1715C}.\footnote{Herschel is an ESA space
  observatory with science instruments provided by European-led Principal Investigator
  consortia and with important participation from NASA.}

Here we present resolved images of the 99 Herculis circumbinary disk. This system is
particularly interesting because unlike most debris disk+binary systems, the binary orbit
is well characterised. The combination of a known orbit and resolved disk means we can
compare their (different) inclinations and consider circumbinary particle dynamics and
formation scenarios.

This system is a first step toward building on the binary debris disk study of
\citet{2007ApJ...658.1289T}. Their \emph{Spitzer} study found that debris disks are as
common in binary systems as in single systems, but tend not to have separations in the
3-30AU range. However, only some of their systems had detections at multiple wavelengths
to constrain the disk location and none were resolved, making the true dust location
uncertain. Resolved systems such as 99 Her remove this ambiguity, and provide crucial
information on the disk location, stability and dynamics.

This paper is laid out as follows. We first consider the stellar and orbital properties
of the 99 Her system, including the possibility of a third component. Then we consider
the \emph{Herschel} image data and several different models that can explain it. Finally,
we discuss the implications of these models for the formation of the system.

\section{99 Herculis}\label{s:stprop}

The binary 99 Herculis (HD 165908, HIP 88745, GJ 704AB, ADS 11077) contains the 37$^{\rm
  th}$ closest F star primary within the volume limited Unbiased Nearby Stars sample
\citep{2010MNRAS.403.1089P}. The Catalogue of Components of Doubles and Multiple systems
\citep[CCDM J18071+3034,][]{2002yCat.1274....0D} lists three components, but using
Hipparcos proper motions \citet{2010MNRAS.403.1089P} find that the 93'' distant C
component is not bound to the system. The binary pair has been known since 1859, and
consists of an F7V primary orbited by a K4V secondary. The primary is known to be metal
poor with [Fe/H] $\approx -0.4$
\citep[e.g.][]{1996yCat..33140191G,2000MNRAS.316..514A,2007PASJ...59..335T} and has an
age consistent with the main-sequence \citep[$6-10$Gyr from isochrone
fitting,][]{2004A&A...418..989N,2007PASJ...59..335T}.

\subsection{Binary Configuration}\label{ss:binary}

\begin{table}
  \caption{99 Her orbital elements, system mass and 1$\sigma$ uncertainties. The ascending node
    $\Omega$ is measured anti-clockwise from North. The longitude of pericenter is
    measured anti-clockwise from the ascending node, and projected onto the sky plane has
    a position angle of 163$^\circ$ (i.e. is slightly different to 41+116 because the
    orbit is inclined).}\label{tab:elem}
  \begin{tabular}{llll}
    \hline
    Parameter & Symbol (unit) & Value & Uncertainty \\
    \hline
    Semi-major axis & a ('') & 1.06 & $0.02$ \\
    Eccentricity & e & 0.766 & $0.004$ \\
    Inclination & i ($^\circ$) & 39 & $2$ \\
    Ascending node & $\Omega$ ($^\circ$) & 41 & $2$ \\
    Longitude of pericenter & $\omega$ ($^\circ$) & 116 & $2$ \\
    Date of pericenter passage & T (yr) & 1997.62 & $0.05$ \\
    Period & P (yr) & 56.3 & $0.1$ \\
    Total mass & $M_{\rm tot} (M_\odot)$ & 1.4 & $0.1$ \\
    \hline
\end{tabular}  
\end{table}

\begin{figure}
  \begin{center}
    \includegraphics[width=0.5\textwidth]{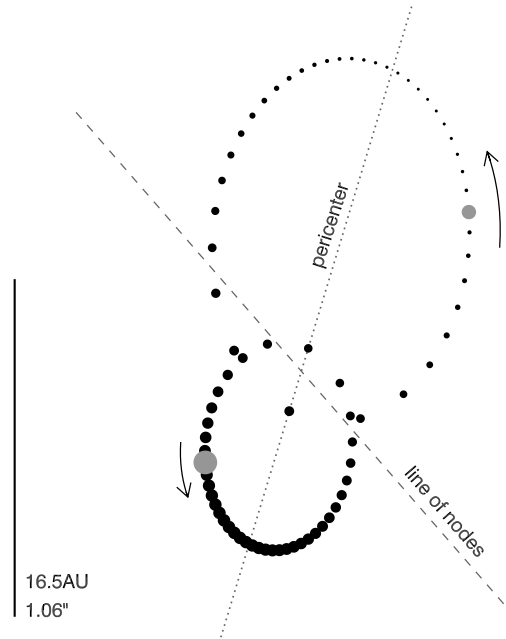}
    \caption{99 Her binary orbit as seen on the sky, with the line of nodes and
      pericenter indicated. North is up and East is left. The stellar orbits are shown
      over one (anti-clockwise) orbital period with black dots. Grey dots (primary is the
      larger grey dot) show the positions at the PACS observation epoch. Black dot sizes
      are scaled in an arbitrary way such that larger dots are closer to Earth. The
      arrows indicate the direction of motion and the scale bar indicates the binary
      semi-major axis of 1.06'' (16.5AU).}\label{fig:sys}
  \end{center}
\end{figure}

To interpret the \emph{Herschel} observations requires an understanding of the binary
configuration, which we address first. Typically, the important orbital elements in
fitting binary orbits are the semi-major axis, eccentricity, and period, which yield
physical characteristics of the system (if the distance is known). Regular observations
of 99 Her date back to 1859 and the binary has completed nearly three revolutions since
being discovered. Additional observations since the previous orbital derivation
\citep{1999A&A...341..121S}, have allowed us to derive an updated orbit. Aside
  from a change of 180$^\circ$ in the ascending node \citep[based on spectroscopic
  data][]{2006ApJS..162..207A}, the orbital parameters have changed little; the main
  purpose of re-deriving the orbit is to quantify the uncertainties.

The orbit was derived by fitting position angles (PAs) and separations ($\rho$) from the
Washington Double Star catalogue
\citep{2011yCat....102026M}.\footnote{\href{http://ad.usno.navy.mil/wds/}{http://ad.usno.navy.mil/wds/}}
We included three additional observations; a \emph{Hubble Space Telescope} (HST)
\emph{Imaging Spectrograph} (STIS) acquisition image \citep[epoch 2000.84, $PA=264 \pm
2^\circ$, $\rho=0.54 \pm 0.02$'',][]{2004ApJ...606..306B}, an adaptive optics image taken
using the Lick Observatory Shane 3m telescope with the IRCAL near-IR camera as an ongoing
search for faint companions of stars in the UNS sample (epoch 2009.41, $PA=309 \pm
2^\circ$, $\rho=1.12 \pm 0.02$''), and a Keck II NIRC2 L' speckle image taken to look for
a third companion (see \S\ref{s:third}, epoch 2011.57, $PA=317 \pm 1^\circ$, $\rho=1.20
\pm 0.014$''). For visual data we set uncertainties of 7$^\circ$ to PAs and 0.5'' to
separations, for \emph{Hipparcos} data we used 5$^\circ$ and 0.1'', and for speckle
observations without quoted uncertainties we used 2$^\circ$ and 0.04''. The resulting
orbital elements, shown in Table \ref{tab:elem}, vary only slightly from those derived by
\citet{1999A&A...341..121S}.\footnote{A figure showing the \citet{1999A&A...341..121S}
  orbit is available in the WDS catalogue.} The best fit yields $\chi^2 = 190$ with 399
degrees of freedom. The fit is therefore reasonable, and most likely the $\chi^2$ per
degrees of freedom is less than unity because the uncertainties assigned to the visual
data are too conservative. If anything, the uncertainties quoted in Table \ref{tab:elem}
are therefore overestimated. However, visual data can have unknown systematic
uncertainties due to individual observers and their equipment so we do not modify them
\citep{2001AJ....122.3472H}.

These data also allow us to derive a total system mass of 1.4$M_\odot$, where we have
used a distance of 15.64pc \citep{2008yCat.1311....0F}. While the total mass is well
constrained by visual observations, the mass ratio must be derived from either the
differential luminosity of each star or radial velocities. We use the spectroscopic mass
function derived by \citet{2006ApJS..162..207A}, yielding a mass ratio of 0.49, which has
an uncertainty of about 10\%. The mass ratio from the differential luminosity is 0.58,
with a much larger (20\%) uncertainty \citep{1999A&A...341..121S}. Using the
spectroscopic result, the primary (GJ 704A) has a mass of 0.94$M_\odot$, while the
secondary (GJ 704B) is 0.46$M_\odot$.

The system configuration is shown in Figure \ref{fig:sys} and is inclined such that the
primary is currently closer to Earth than the secondary. The position of the B component
relative to A on the date it was observed by \emph{Herschel} in late April 2010 was
PA=314$^\circ$ at an observed separation of 1.15'' (22.6AU when deprojected), indicated
by grey circles in the Figure.

\subsection{A Third Component?}\label{s:third}

\begin{figure}
  \begin{center}
    \hspace{-0.5cm} \includegraphics[width=0.5\textwidth]{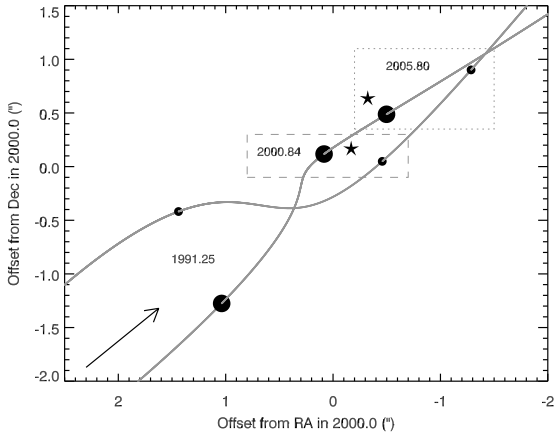}
    \caption{Motion of the 99 Her system (filled dots) in sky coordinates at three
      epochs. The epochs including the putative third component are enclosed in
      boxes. The arrow shows the direction of the system center of mass movement and the
      distance travelled in 5 years, and the grey lines show the path traced out by each
      star. Star symbols show the position of the third object observed in the STIS data
      in 2000 (dashed box) and by \citet{2008AN....329...54S} in 2005 (dotted
      box).}\label{fig:pm}
  \end{center}
\end{figure}

While the STIS images clearly resolve the binary, there is a possible third component
with $PA \approx 284^\circ$ and $\rho \approx 0.27"$ that is about 2.4 times as faint as
the B component. \citet{2008AN....329...54S} also report a third component (epoch 2005.8)
at $PA \approx 50^\circ$ and $\rho \approx 0.228"$ (no magnitude is given). However,
while they detected the secondary again in mid-2007, they do not report any detection of
the putative tertiary \citep{2010AN....331..286S}. The detected positions are shown as
star symbols in sky coordinates in Figure \ref{fig:pm}, which shows the motion of the 99
Her system. The system proper motion is $\mu_\alpha \cos \delta = -110.32$mas yr$^{-1}$,
$\mu_\delta = 110.08$mas yr$^{-1}$ \citep{2007ASSL..350.....V}, and accounts for the
motion of the primary assuming the orbit derived in \citet{1999A&A...341..121S}, which is
very similar to ours.  The small proper motion uncertainty means STIS and
\citet{2008AN....329...54S} cannot have seen the same object if it is fixed on the
sky. There is no clear sign of a third component in the residuals from fitting the orbit
of the secondary.


\begin{figure}
  \begin{center}
    \includegraphics[width=0.45\textwidth]{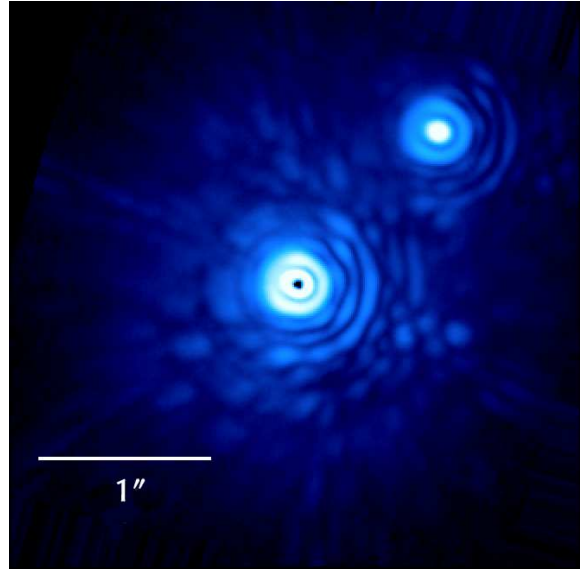}
    \caption{Keck/NIRC2 adaptive optics image of 99 Her at 3.8$\mu$m, cropped to about
      1.5'' around the primary. North is up and East is left. The saturated A component
      is at the center of the frame.}\label{fig:keck}
  \end{center}
\end{figure}  

To try and resolve this issue we obtained an adaptive optics image of 99 Her at L' (3.8
$\mu$m) using the NIRC2 camera at Keck II on July 27, 2011, shown in Figure
\ref{fig:keck}.  We adopted the narrow camera (10 mas/pixel) and used a five-point dither
pattern with three images obtained at each position consisting of 25 coadds of 0.181
seconds integration.  The cumulative integration time for the final co-registered and
coadded image is 67.875 seconds. The core of the A component point-spread-function is
highly saturated, which degrades the achievable astrometry. We estimate the position of
99 Her A by determining the geometric center of the first diffraction ring. The position
of 99 Her B is taken from the centroid of the unsaturated core. The PA and separation are
quoted above.

There is no detection of the putative 99 Her C within 1.6'' of the primary in the
combined image if it is only a factor 2.4 fainter than the B component, because it would
appear 20 times brighter than the brightest speckles. However, if it were closer to the
primary than 0.2'' it would currently be too close to detect. If the object was fixed on
the sky near either the 2000 or 2005 locations, it would have been detected in the
individual pointings of the five-point dither since each NIRC2 pointing has a field of
view of $10''\times10''$. To be outside the field of view and still bound to the primary,
the tertiary must have an apocenter larger than about 75AU (5''). An object in such an
orbit would have a period of at least 200 years, so could not have been detected near the
star in 2005 and be outside the NIRC2 field of view in 2011.

The non-detections by \citet{2010AN....331..286S} and NIRC2 make the existence of the
tertiary suspicious. It is implausible that the object was too close to, or behind the
star both in 2007 and 2011, because at a semi-major axis of 0.23'' (3.5AU) from the
primary (similar to the projected separation) the orbital period is 7 years. Therefore,
the object would be on opposite sides of the primary, and the two detections already rule
out an edge-on orbit. Even assuming a circular orbit, such an object is unlikely to be
dynamically stable, given the high eccentricity and small pericenter distance (4.1AU) of
the known companion. A tertiary at this separation would be subject to both short term
perturbations and possible close encounters. If the mutual inclination were high enough,
it would also be subject to Kozai cycles \citep{1962P&SS....9..719L,1962AJ.....67..591K}
due to the secondary that could result in a high eccentricity and further affect the
orbital stability.

While it may be worthy of further detection attempts, the existence of this component
appears doubtful and we do not consider it further.

\section{IR and Sub-mm Data}\label{s:data}

\subsection{Observations}\label{s:obs}

\begin{figure*}
  \begin{center}
    \hspace{-0.5cm} \includegraphics[width=0.33\textwidth]{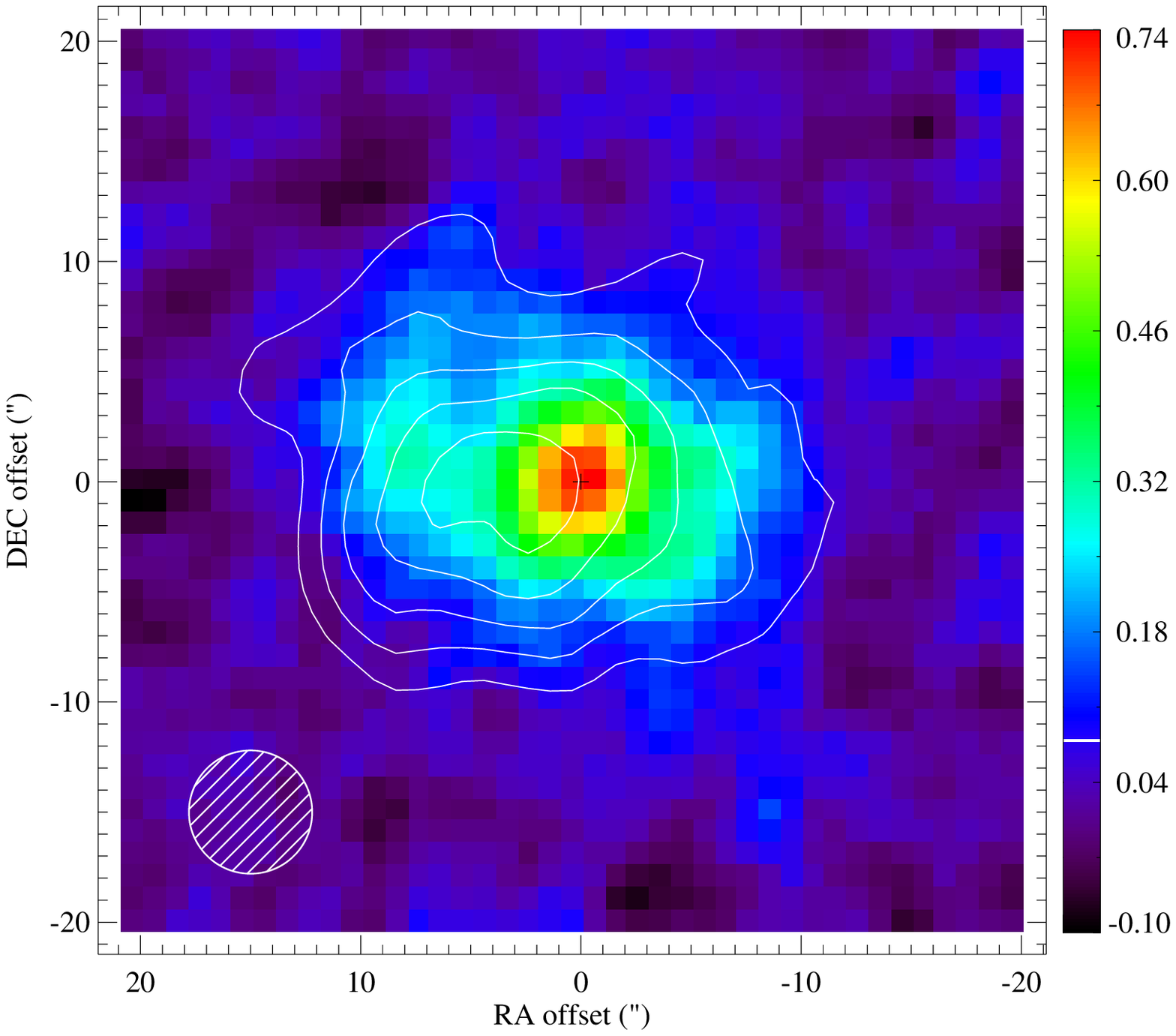} 
         \includegraphics[width=0.33\textwidth]{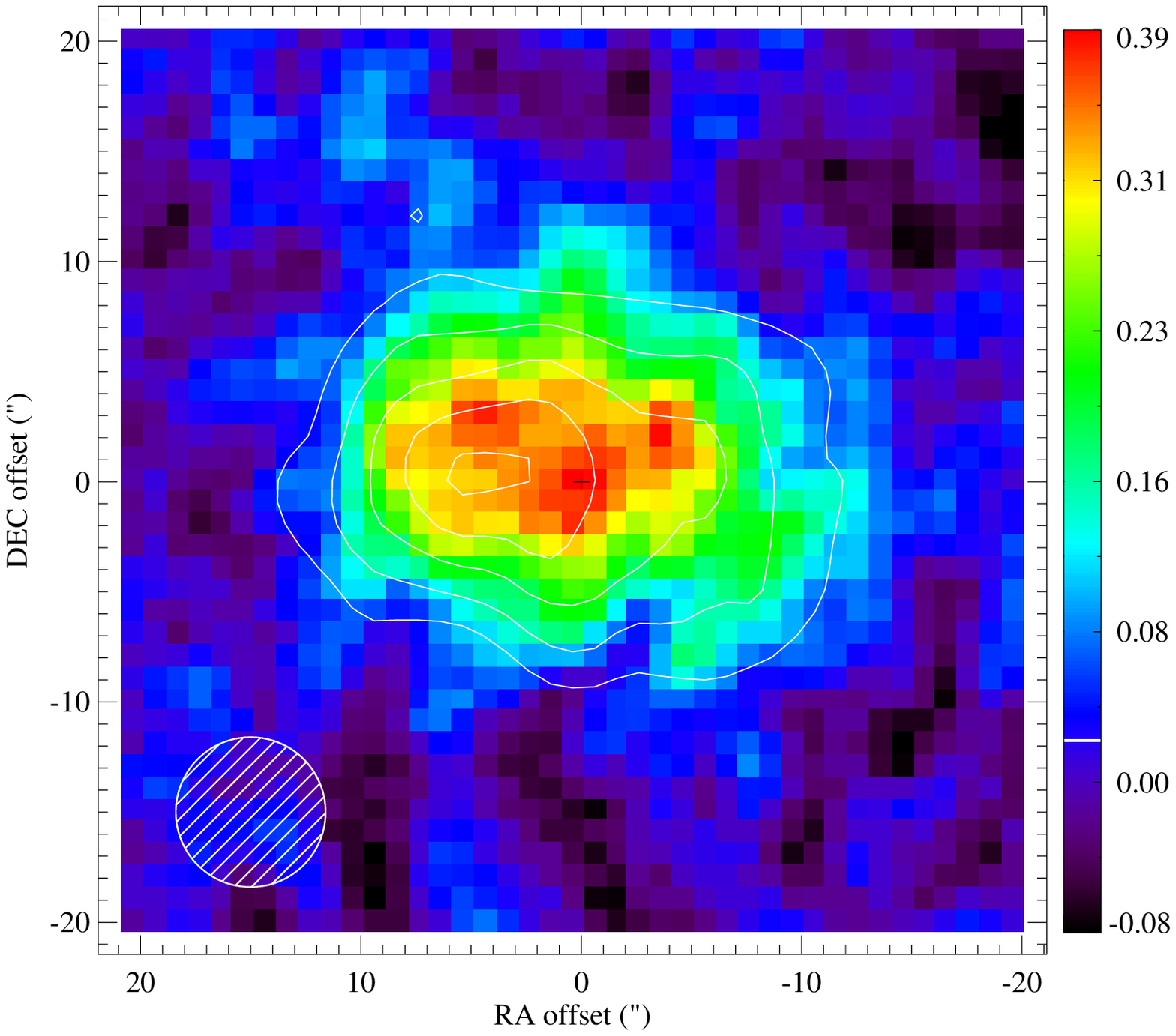}
         \includegraphics[width=0.33\textwidth]{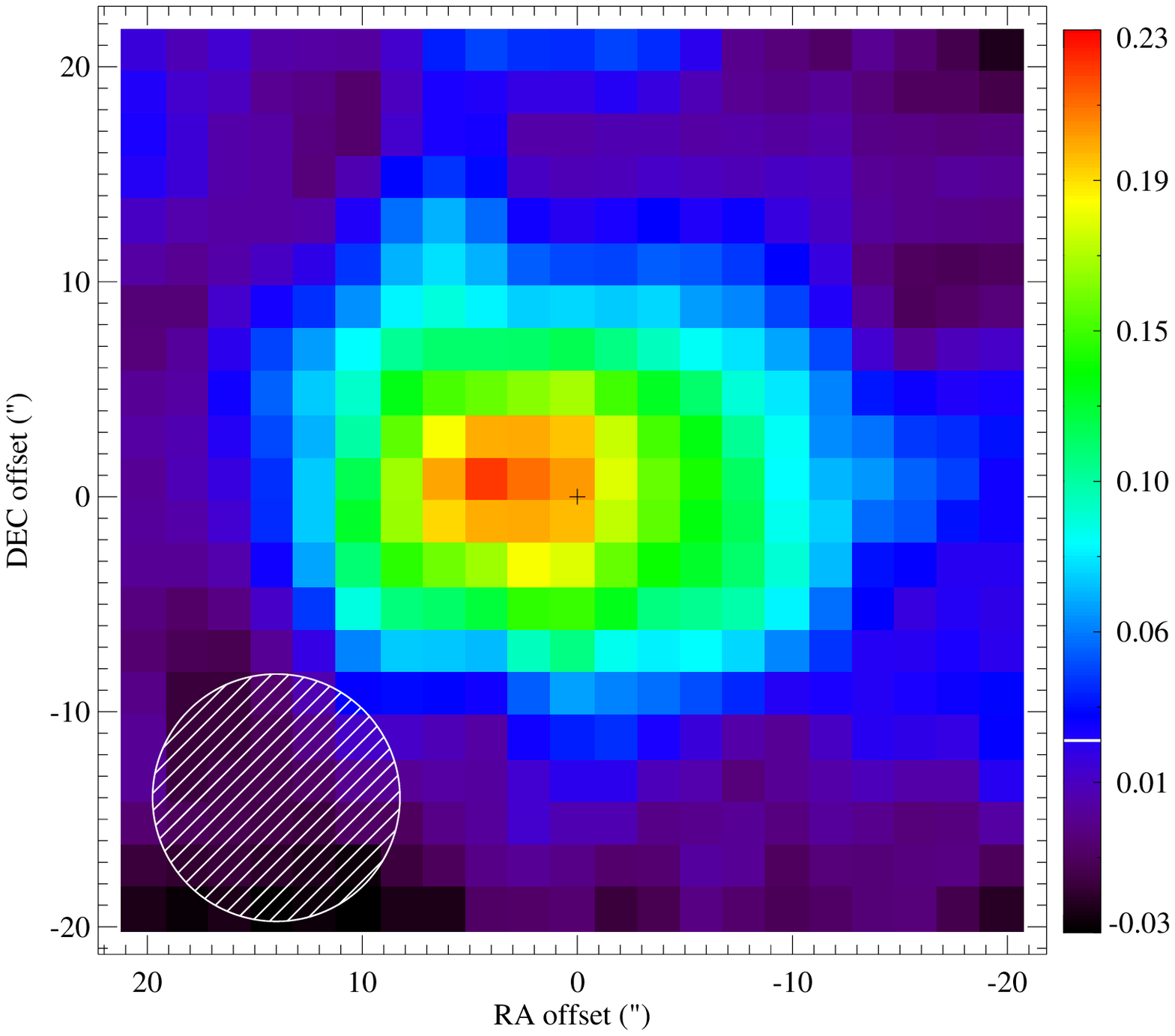}
         \caption{\emph{Herschel} PACS images of the 99 Her system at 70$\mu$m (left)
           100$\mu$m (center) and combined 160$\mu$m (right). North is up and East is
           left. Colour bars show levels in units of mJy/arcsec$^2$. Contours in the left
           two panels show the 160$\mu$m data from each observation in 5 levels from
           3-9$\sigma$. The images are centered on the primary position (plus symbol)
           derived from the 70$\mu$m image. Hatched circles show the FWHM beam sizes of
           5.6'', 6.8'', and 11.5'' at 70, 100, and 160$\mu$m
           respectively.}\label{fig:pacs}
  \end{center}
\end{figure*}

\emph{Herschel} Photodetector and Array Camera \& Spectrometer
\citep[PACS,][]{2010A&A...518L...2P} data at 100 and 160$\mu$m were taken in April 2010
during routine DEBRIS observations. Subsequently, a Spectral \& Photometric Imaging
Receiver \citep[SPIRE,][]{2010A&A...518L...3G} observation was triggered by the large
PACS excess indicating the presence of a debris disk and a likely sub-mm detection. The
disk was detected, but not resolved with SPIRE at 250 and 350$\mu$m. A 70$\mu$m PACS
image was later obtained to better resolve the disk. Because every PACS observation
includes the 160$\mu$m band, we have two images at this wavelength, which are combined to
produce a single higher S/N image. All observations were taken in the standard scan-map
modes for our survey; mini scan-maps for PACS data and small maps for SPIRE. Data were
reduced using a near-standard pipeline with the Herschel Interactive Processing
Environment \citep[HIPE Version 7.0,][]{2010ASPC..434..139O}. We decrease the noise
slightly by including some data taken as the telescope is accelerating and decelerating
at the start and end of each scan leg.

The high level of redundancy provided by PACS scan maps means that the pixel size used to
generate maps can be smaller than the natural scale of 3.2''/pix at 70 and 100$\mu$m and
6.4''/pix at 160$\mu$m via an implementation of the ``drizzle'' method
\citep{2002PASP..114..144F}. Our maps are generated at 1''/pix at 70 and 100$\mu$m and
2''/pix at 160$\mu$m. The benefit of better image sampling comes at the cost of
correlated noise \citep{2002PASP..114..144F}, which we discuss below.

In addition to correlated noise, two characteristics of the PACS instrument combine to
make interpretation of the data challenging. The PACS beam has significant power at large
angular scales; about 10\% of the energy lies beyond 1 arcminute and the beam extends to
about 17 arcminutes (1000 arcsec). While this extent is not a problem in itself, it
becomes problematic because PACS data are subject to fairly strong $1/f$ (low frequency)
noise and must be high-pass filtered. The result is that a source will have a flux that
is 10-20\% too low because the ``wings'' of the source were filtered out. While this
problem can be circumvented with aperture photometry using the appropriate aperture
corrections derived from the full beam extent, the uncorrected apertures typically used
for extended sources will result in underestimates of the source flux.\footnote{See
  http://herschel.esac.esa.int/twiki/bin/view/Public/WebHome for details regarding the
  PACS beam extent and calibration.}

Here, we correct the fluxes measured in apertures for 99 Her based on a comparison
between PSF fitted and aperture corrected measurement of bright point sources in the
DEBRIS survey with predictions from their stellar models \citep[based on the calibration
of ][]{2008AJ....135.2245R}. These upward corrections are $16 \pm 5\%$, $19 \pm 5\%$, and
$21 \pm 5\%$ at 70, 100, and 160$\mu$m respectively. These factors depend somewhat on the
specifics of the data reduction, so are \emph{not} universal. This method assumes that
the correction for 99 Her is the same as for a point source, which is reasonable because
the scale at which flux is lost due to filtering the large beam is much larger than the
source extent. The corrected PACS measurement is consistent with MIPS 70$\mu$m, so we do
not investigate this issue further.

The beam extent and filtering is also important for resolved modelling because the
stellar photospheric contribution to the image is decreased. Therefore, in generating a
synthetic star+disk image to compare with a raw PACS observation, the stellar
photospheric flux should be decreased by the appropriate factor noted
above. Alternatively, the PACS image could be scaled up by the appropriate factor and the
true photospheric flux used.

Table \ref{tab:obs} shows the measured star+disk flux density in each \emph{Herschel}
image. Uncertainties for PACS are derived empirically by measuring the standard deviation
of the same sized apertures placed at random image locations with similar integration
time to the center (i.e. regions with a similar noise level).

The SPIRE observations of 99 Her are unresolved. The disk is detected with reasonable S/N
at 250$\mu$m, marginally detected at 350$\mu$m, and not detected at 500$\mu$m. Fluxes are
extracted with PSF fitting to minimise the contribution of background objects. Because
all three bands are observed simultaneously (i.e. a single pointing), the PSF fitting
implementation fits all three bands at once. A detection in at least one band means that
all fluxes (or upper limits) are derived at the same sky position.

Additional IR data exist for 99 Her, taken with the Multiband Imaging Photometer for
\emph{Spitzer} \citep[MIPS,][]{2004ApJS..154...25R}. Only the star was detected at
24$\mu$m ($270.3 \pm 0.1$mJy), but this observation provides confirmation of the 99 Her
stellar position in the PACS images relative to a background object 1.8 arcmin away to
the SE ($PA=120^\circ$) that is visible at 24, 70, and 100$\mu$m. The presence of an
excess at 70$\mu$m ($98 \pm 5$mJy compared to the photospheric value of 30mJy) was in
fact reported by \citet{2010ApJ...710L..26K}. They did not note either the circumbinary
nature or that the disk may be marginally resolved by MIPS at 70$\mu$m. Because our study
focuses on the spatial structure, we use the higher resolution PACS data at 70$\mu$m, but
include the MIPS data to model the SED.

\subsection{Basic image analysis}\label{s:basic}

Figure \ref{fig:pacs} shows the \emph{Herschel} PACS data. Compared to the beam size, the
disk is clearly resolved at all three wavelengths. At 160$\mu$m the peak is offset about
5'' East relative to both the 70 and 100$\mu$m images. However, the disk is still visible
at 160$\mu$m as the lower contours match the 70 and 100$\mu$m images well. The 160$\mu$m
peak is only 2-3$\sigma$ more significant than these contours. While such variations are
possible due to noise, in this case the offset is the same in both 160$\mu$m images, so
could be real. The fact that the peak extends over several pixels is not evidence that it
is real, because the pixels in these maps are correlated (see below). If real, this
component of the disk or background object cannot be very bright at SPIRE wavelengths
because the measured fluxes appear consistent with a blackbody fit to the disk (see
\S\ref{s:sed}). Based on an analysis of all DEBRIS maps (that have a constant depth), the
chance of a 3$\sigma$ or brighter background source appearing within 10'' of 99 Her at
160$\mu$m is about 5\% (Thureau et al in prep). Given that the 160$\mu$m offset is only a
2-3$\sigma$ effect (i.e. could be a 2-3$\sigma$ background source superimposed on a
smooth disk), the probability is actually higher because the number of background sources
increases strongly with depth. These objects have typical temperatures of 20--40K
\citep[e.g.][]{2010A&A...518L...9A}, so could easily appear in only the 160$\mu$m image,
particularly if the disk flux is decreasing at this wavelength.

\begin{table}
  \caption{\emph{Herschel} photometry of 99 Her. The disk is not detected at 500$\mu$m
    and can be considered a 3$\sigma$ upper limit of 24mJy.}\label{tab:obs}
  \begin{tabular}{lrrl}
    Band & Flux (mJy) & Uncertainty & Method \\
    \hline
    PACS70 & 93 & 10 & 15'' aperture \\
    PACS100 & 87 & 10 & 15'' aperture \\
    PACS160 & 80 & 15 & 17'' aperture \\
    SPIRE250 & 44 & 6 & PSF fit \\
    SPIRE350 & 22 & 7 & PSF fit \\
    SPIRE500 & 4 & 8 & PSF fit \\
  \end{tabular}  
\end{table}

We now analyse the PACS images using 2D Gaussian models to estimate the disk size,
inclination, and position angle. A 2D Gaussian fits the star-subtracted PACS 100$\mu$m
image fairly well, with major and minor full-width half-maxima of 17.7 and 12.8'' at a
position angle of $78^\circ$. Quadratically deconvolving from the 6.7'' FWHM beam
assuming a circular ring implies an inclination of $48^\circ$ from face-on and an
estimated diameter of 250AU. Gaussian fitting to star-subtracted images at both 70 and
160$\mu$m yields similar results.

As noted above, estimation of uncertainties in these parameters is non-trivial due to
correlated noise, but made easier by the constant depth of our survey. By inserting the
best fit Gaussian to the star-subtracted image of the 99 Her disk from the 100$\mu$m
image into 438 other 100$\mu$m maps at high coverage positions offset from the intended
target, we obtain a range of fit parameters for hundreds of different realisations of the
same noise. This process yields an inclination of $45 \pm 5^\circ$ and PA of $75 \pm
8^\circ$. Repeating the process, but using the best fit Gaussian for the 70$\mu$m image
yields an inclination of $44 \pm 6^\circ$ and PA of $68 \pm 9^\circ$. Though the
inclination of the disk is similar to the binary, the position angle is significantly
different from the binary line of nodes of $41 \pm 2 ^\circ$. This difference means that
the disk and binary orbital planes are misaligned.

As a check on the above approach, we can correct for the correlated noise
directly. \citet{2002PASP..114..144F} show that for a map that has sufficiently many
dithers (corresponding in our case to many timeline samples across each pixel), a noise
``correction'' factor of $r/\left(1-1/3r\right)$ can be derived, where $r$ is the ratio
of natural to actual pixel scales and is 3.2 for our PACS maps. A correction factor of
3.6 for the measured pixel to pixel noise is therefore required when estimating the
uncertainty on a fitted Gaussian. Including this factor at 70$\mu$m and calculating the
uncertainty by the standard $\Delta \chi^2$ method yields an inclination of $42 \pm
7^\circ$ and a PA of $68 \pm 9^\circ$. At 100$\mu$m the result is an inclination of $44
\pm 6^\circ$ and a PA of $76 \pm 8^\circ$. These results are therefore almost exactly the
same as the empirical method used above and therefore lead to the same conclusion of
misalignment.

As will become apparent in \S\ref{s:spatial}, there is reason to believe that the disk
plane could be perpendicular to the binary pericenter direction. The projection of the
binary pericenter direction on the sky plane has a PA of $163 \pm 2^\circ$, and a line
perpendicular to this has a PA of $73 \pm 2^\circ$. Therefore, the observed disk position
angle of about 72$^\circ$ is consistent with being at 90$^\circ$ to the binary pericenter
direction.


\section{SED}\label{s:sed}

\begin{figure}
  \begin{center}
    \hspace{-0.5cm} \includegraphics[width=0.5\textwidth]{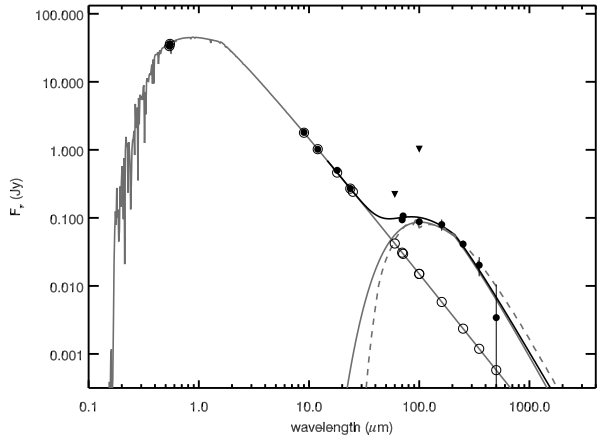}
    \caption{SED for the 99 Her system (both stars) showing the stellar and disk models
      (grey lines) and star+disk model (black line). The blackbody disk model is the
      solid grey line, and the physical grain model the dashed line. Photometric
      measurements are shown as black filled circles, and synthetic photometry of the
      stellar atmosphere as open circles ($U-B$, $B-V$, \& $b-y$ colours, and $m1$ and
      $c1$ Stromgren indices were fitted but are not shown here). Black triangles mark
      upper limits from IRAS at 60 and 100$\mu$m.}\label{fig:sed}
  \end{center}
\end{figure}  

The combination of all photometry for 99 Her allows modelling of the spectral energy
distribution (SED). The model is separated into two components; a stellar atmosphere and
a disk. Due to being fairly bright ($V\sim5$mag) the system is saturated in the 2MASS
catalogue. However, sufficient optical photometry for each individual star and the pair
exists
\citep{1993AJ....106..773H,1997yCat.2215....0H,1997ESASP1200.....P,2006yCat.2168....0M},
as well as infra-red measurements of the AB pair from AKARI and IRAS
\citep{1990IRASF.C......0M,2010A&A...514A...1I}. These data were used to find the best
fitting stellar models via $\chi^2$ minimisation. This method uses synthetic photometry
over known bandpasses and has been validated against high S/N MIPS 24$\mu$m data for
DEBRIS targets, showing that the photospheric fluxes are accurate to a few percent for
AFG-type stars. The stellar luminosities ($L_{\star,A}=1.96L_\odot$,
$L_{\star,B}=0.14L_\odot$) and IR fluxes of the individual components are consistent with
the fit for the pair ($L_{\star,AB}=2.08L_\odot$). The fit for the AB pair is shown in
Figure \ref{fig:sed}.

The spatial structure of the disk can be modelled with dust at a single radial distance
of 120AU (i.e. thin compared to \emph{Herschel's} resolution, \S\ref{s:spatial}), so disk
SED modelling can be decoupled from the resolved modelling once this radial distance is
known. Because we have measurements of the disk emission at only five wavelengths, we
cannot strongly constrain the grain properties and size distribution. We fit the data
with a blackbody model, and then compare the data with several ``realistic'' grain models
\citep{1997A&A...323..566L,2001A&A...370..447A,2002MNRAS.334..589W}.

In fitting a blackbody we account for inefficient grain emission at long wavelengths by
including parameters $\lambda_0$ and $\beta$, where the blackbody is modified by a factor
$\left( \lambda_0/\lambda \right)^\beta$ for wavelengths longer than $\lambda_0$. The
best fitting model has a temperature of 49K and fractional luminosity $L_{\rm
  disk}/L_\star = 1.4 \times 10^{-5}$. The SPIRE data are near the confusion limit of
about 6mJy, so the parameters $\beta$ and $\lambda_0$ are unconstrained within reasonable
limits by the data (based on previous sub-mm detections for other disks we fix them to
$\lambda_0=210 \mu$m and $\beta=1$ in Figure \ref{fig:sed} \citep{2007ApJ...663..365W}).

Assuming that grains absorb and emit like blackbodies, the radial dust distance implied
by 49K is 45AU. Because the disk is observed at a radius of 120AU (i.e. is warmer than
expected for blackbodies at 120AU), the dust emission has at least some contribution from
grains small enough to emit inefficiently in the 70-350$\mu$m wavelength range. Because
the SED alone is consistent with a pure blackbody (i.e. with $\beta=0$), we cannot make
such a statement without the resolved images. However, actually constraining the grain
sizes is difficult because temperature and emission are also affected by composition. We
fit the data by generating disk SEDs for grains at a range of semi-major axes and
choosing the one with the lowest $\chi^2$. If the dust semi-major axis is different from
the observed semi-major axis of 120AU the model parameters are changed and the model
recalculated, thus iterating towards the best fit.

We model the dust with a standard diameter ($D$) distribution $n(D) \propto D^{2-3q}$
where $q=1.9$ \citep[equivalently $n(M) \propto M^{-q}$ where $M$ is
mass][]{2003Icar..164..334O}, with the minimum size set by the blowout limit for the
specific composition used (about 1.4$\mu$m) and a maximum size of 10cm. The size
distribution probably extends beyond 10cm, but objects larger than 10cm contribute
negligibly to the emission because the size distribution slope means that smaller grains
dominate. Preliminary tests found that icy grains provided a much better fit than
silicates. To refine the grain model so the SED and resolved radius agree, we introduced
small amounts of amorphous silicates to the initially icy model. The grains are therefore
essentially ice mixed with a small fraction ($f_{\rm sil} = 1.5\%$) of silicate. The icy
grain model is shown as a dotted line in Figure \ref{fig:sed}. This model has a total
dust surface area of 14AU$^2$ and a mass of order 10$M_\oplus$ if the size distribution
is extrapolated up to 1000km size objects.

The parameters from of this model are degenerate for the data in hand; for example the
size distribution could be shallower and the fraction of silicates higher (e.g. $q=1.83$
and $f_{\rm sil} = 4\%$). If we allow the minimum grain size to be larger than the
blowout limit, the disk is well fit by amorphous silicate grains with $q=1.9$ and $D_{\rm
  bl} = 10\mu$m. The disk spectrum can even be fit with a single size population of
25$\mu$m icy grains. However, the predictions for the flux at millimeter wavelengths
depend on the size distribution, with lower fluxes for steeper size
distributions. Therefore, grain properties and size distribution can be further
constrained in the future with deep (sub)mm continuum measurements.

In summary, it is hard to constrain the grain sizes or properties. There is a difference
in the required minimum grain size that depends on composition. Because icy grains are
reflective at optical wavelengths, a detection of the disk in scattered light could
constrain the albedo of particles, and therefore their composition.

\section{Spatial structure}\label{s:spatial}

\begin{figure*}
  \begin{center}
    \hspace{-0.5cm} \includegraphics[width=1\textwidth]{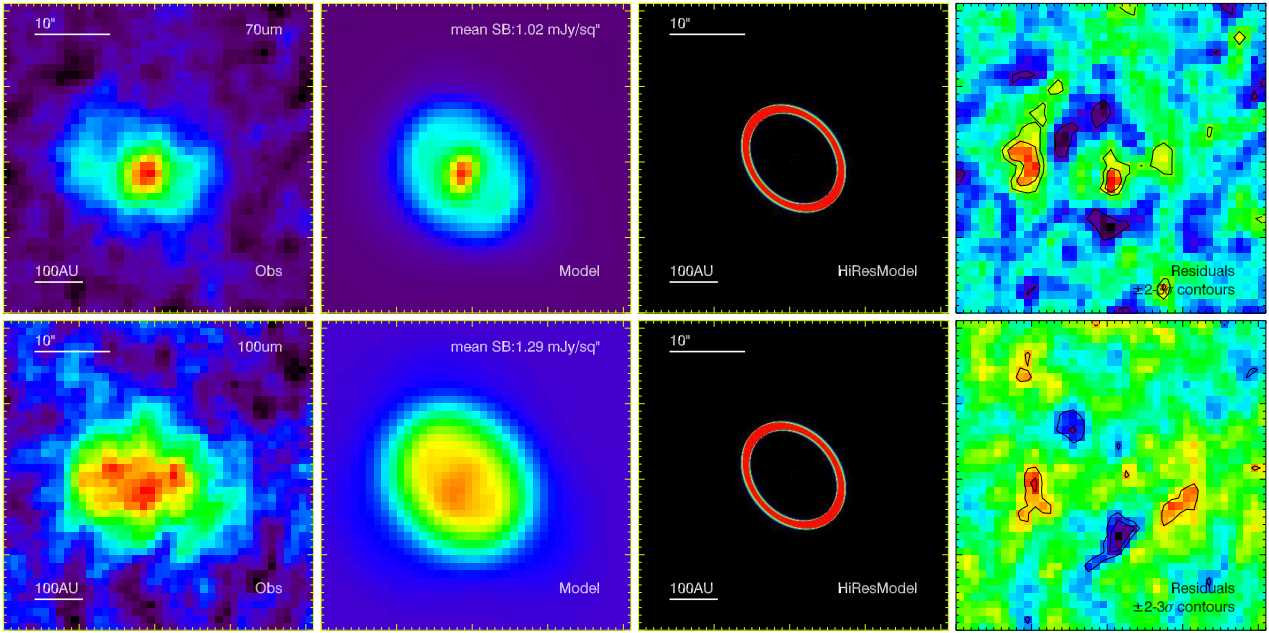}
    \caption{Simple ring model where the disk is aligned with the binary orbital plane,
      compared to the PACS 70 (top row) and 100$\mu$m (bottom row) observations. North is
      up and East is left. Columns show the data, model convolved with the PACS beam, a
      high resolution model, and the data-model residuals from left to right. The model
      includes stellar photospheric fluxes, which lie in two pixels near the ring center
      but are not visible in the high resolution plot. In the residual plots, contours
      show $\pm$2 and 3$\sigma$ in units of pixel to pixel RMS.}\label{fig:copl}
  \end{center}
\end{figure*}  

The PACS images of the 99 Her disk are resolved, which allows modelling of the spatial
distribution of grains that contribute to the observed emission at each wavelength. We
compare synthetic images with the \emph{Herschel} observations in several steps: i)
Generate a three dimensional distribution of surface area $\sigma(r,\theta,\phi)$, where
the coordinates are centered on the primary star. ii) Generate a radial distribution of
grain emission properties. Because the SED can be modelled with blackbody grains at 49K
and the spatial structure modelled with a narrow ring, there is no real need for a radial
temperature dependence and the grain properties are only a function of wavelength:
$P(\lambda) = B_\nu(49K,\lambda)$. Practically, we use a radial temperature dependence $T
\propto r^{-1/2}$ centered on the primary, normalised so that the disk ring temperature
is 49K. This approach ensures that temperature differences due to non-axisymmetries
(negligible for 99 Her) are automatically taken into account. iii) Generate a high
resolution model as viewed from a specific direction. The emission in a single pixel of
angular size $x$ from a single volume element in the three dimensional model $dV$ viewed
along a vector $\mathcal{R}$ is $dF_\nu(\lambda,r,\theta,\phi) = P(\lambda)
\sigma(r,\theta,\phi) dV$, where $dV=x^2 d^2 d\mathcal{R}$, so $d\mathcal{R}$ is the
length of the volume element, and $d$ is the distance to the particles from Earth
\citep{1999ApJ...527..918W}. The emission is derived by integrating along the line of
sight $\mathcal{R}$ for each pixel in the synthetic image. The photospheric fluxes for
each star (decreased by the factors noted in \S\ref{s:obs}) are placed in the relevant
pixels at this step. iv) Convolve the high resolution model with a high resolution
telescope+instrument beam, for which we use interpolated and rotated PACS images of the
star Arcturus.\footnote{The observations are reduced in sky coordinates, so the
  spacecraft orientation is different for each observation and the PSF must be rotated
  accordingly. This rotation step could be avoided by reducing both in spacecraft
  coordinates.} v) Degrade the resolution to match the data. vi) Generate a map of
residuals, defined by $(observed-model)/uncertainty$, where the uncertainty is the pixel
to pixel RMS for that observation. We compute the model $\chi^2$ from pixels in a square
crop around the disk.

A minor consideration is that in the general circumbinary case the disk temperature is
not axisymmetric because the disk orbits the center of mass, not the primary. An
axisymmetric disk is therefore subject to a temperature asymmetry such that it will be
slightly hotter, and therefore brighter, where the distance to the primary is
smallest. This ``binary offset'' asymmetry will rotate with the primary, and will be most
pronounced in the coplanar case. The result of this effect is similar to the offset
caused by perturbations from an eccentric object \citep[`` pericenter
glow''][]{1999ApJ...527..918W}. However, the pericenter glow is offset towards the
pericenter of the perturbing object, so does not rotate unless the perturbing object's
pericenter precesses. The offset from the primary and the pericenter glow are completely
independent effects. Therefore, if the pericenter glow effect is present, it will
reinforce and cancel the binary offset effect, depending on the relative magnitude of
each each offset. The magnitude of the binary offset effect is negligibly small
($\lesssim$1\%) because the disk radius is much larger than the binary
separation. Because our model is centered on the system center of mass this effect is
taken into account anyway. We discuss the effect of the binary on pericenter glow in
\S\ref{s:dyn}.
 
To fit the data requires a handful of parameters, some are required for all models and
some are model specific. The disk surface area, temperature, radius, width, and total
opening angle are the five physical parameters for a ring. The sky position angle and
inclination are two further parameters that set the orientation, but can be fixed if the
disk plane is assumed to be aligned with the binary. In addition each observation has the
stellar RA and Dec as parameters to allow for the 2'' $1\sigma$ pointing accuracy of
\emph{Herschel}. The position at 160$\mu$m is tied to the 100$\mu$m position. There are
therefore eleven possible parameters to fit for the resolved observations at 70, 100, and
160$\mu$m. We fix the disk temperature to 49K in all cases.

From the basic analysis (\S\ref{s:basic}), a simple ring coplanar with the binary does
not appear a viable option. To emphasise this point we show 70 and 100$\mu$m images of
the best fitting coplanar model in Figure \ref{fig:copl}. This model was generated by the
steps outlined above, and the rightmost three panels are the results of steps iv
(convolved model), iii (high resolution model), and vi (residuals). We fix the disk width
to 20AU, the opening angle to 5$^\circ$, and the position angle and inclination to the
binary plane, so there are six free parameters (surface area, radius, and two pairs of
RA/Dec sky positions). While we include the 160$\mu$m data in the fitting, it does not
constrain the fit strongly due to low S/N and always shows $\sim$2$\sigma$ residual
structure due to the offset peak. For comparison with the models below, the $\chi^2$
value for all three PACS bands is 4278 with 3797 degrees of freedom. The positive and
negative residuals (rightmost panels) show that the disk ansae in the model have the
wrong orientation at both wavelengths. It is clear that any structure symmetric about the
binary line of nodes will not be consistent with the observations because the position
angle is significantly different.

An alternative explanation for the misalignment between the observed position angle and
the binary line of nodes could be that the dust does in fact lie in the binary plane, but
that the particles are on eccentric orbits with common pericenter directions (i.e. the
disk is elliptical and offset from the binary). In principle, the observations can
constrain the eccentricity and pericenter direction. However, this model fails because
the eccentricity needed to match the observed position angle is too extreme. In order to
obtain an ellipse that lies in the binary orbital plane and has a position angle and
aspect ratio similar to the observations requires eccentricities $\gtrsim$0.4. The
eccentricity of these particles is so high that i) the ring is significantly offset from
the star and ii) the ring has an extreme pericenter glow asymmetry at all wavelengths
caused by particles residing at different stellocentric distances. Because the PACS
70$\mu$m image shows that the star lies very near the disk center, such a strong offset
is ruled out.

We now consider two relatively simple models that account for the misalignment between
the disk and binary orbital planes. The first is based on the expected secular evolution
of circumbinary particles, and the second is a simple misaligned ring where the disk
position angle and inclination are free parameters.

\subsection{Secularly perturbed polar ring}\label{s:polar}

In this section we consider a ring inspired by the secular evolution of circumbinary
particles. This approach ensures that the disk is stable over the stellar lifetime and
encompasses the particle dynamics dictated by the binary. We first outline the dynamics
of circumbinary particles, and then show the model for the 99 Her disk.

\subsubsection{Dynamics}\label{s:dyn}

Particle dynamics are important for evolution and stability in the 99 Her system. A
circumbinary disk will have its inner edge truncated, while circumstellar disks around
either component can be truncated at their outer edges. In addition, secular
perturbations lead to precession of test particles' nodes coupled with inclination
variations. We explore these dynamics using the \emph{Swift HJS} integrator
\citep{2003A&A...400.1129B}.

In general, disk truncation allows us to place limits on possible locations for disk
particles. However, in the case of 99 Her there is no evidence for disk components
orbiting only one star, and the apparent circumbinary disk extent lies well beyond
$\sim$30-60AU stability limit at any inclination
\citep{1997AJ....113.1445W,2011arXiv1108.4144D}.

Circumbinary particles also undergo long-term dynamical evolution due to secular
perturbations. Because the binary has a small mass ratio and high eccentricity, the
dynamics are not well described by the circular restricted three-body problem, commonly
applied in the case of debris disks perturbed by planets. Similar dynamics have
previously been explored in the context of the HD 98800 system
\citep{2008MNRAS.390.1377V,2009MNRAS.394.1721V} and more generally
\citep{2010MNRAS.401.1189F,2011arXiv1108.4144D}.

These studies show that the inclination ($i$) and line of nodes ($\Omega$) of
circumbinary particles evolve due to perturbations from the binary. Depending on the
binary eccentricity and particle inclination, $\Omega$ can circulate (increase or
decrease continuously) or librate (oscillate about 90 or $270^\circ$). Particles with low
inclinations stay on low inclination orbits, thus sweeping out a roughly disk or
torus-like volume over long timescales. Higher inclination particles are subject to nodal
libration and large inclination variations, thus sweeping out large parts of a sphere
around the binary. Most importantly for 99 Her, the orbits of particles with $\Omega
\approx 90^\circ$ (or $270^\circ$) and on near polar orbits will not change much due to
secular evolution, thus sweeping out a polar ring.

\begin{figure}
  \begin{center}
    \hspace{-0.5cm} \includegraphics[width=0.5\textwidth]{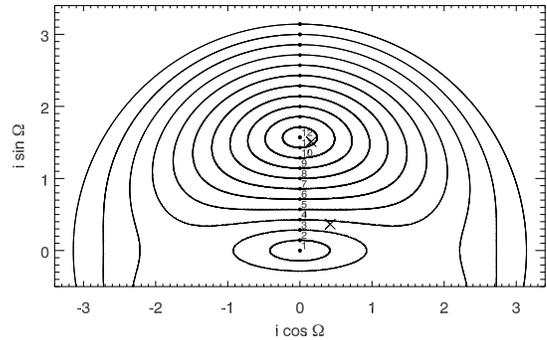}
    \caption{Secular evolution of circumbinary particles in inclination ($i$) and line of
      nodes ($\Omega$) space. Particles begin at dots and move along the curves due to
      perturbations from the binary. Particles that would appear reflected in the x axis
      duplicate the spatial distribution so are not shown. Crosses show the current
      location of particles in the two interpretations of the transient ring model
      (\S\ref{s:ring}). Over time these particles will sweep out curves similar to
      particles 4 and 11. The long term structure of the transient ring will therefore
      appear similar to either panel 4 or 11 in Figure \ref{fig:swift}, depending on
      which inclination is correct.}\label{fig:sec}
  \end{center}
\end{figure}  

Figure \ref{fig:sec} shows the secular evolution of 23 particles on initially circular
orbits in complex inclination space. All particles have initial nodes of 90$^\circ$
relative to the binary pericenter and inclinations spread evenly between 0 and
180$^\circ$ and are integrated for 1Gyr (i.e. there are no other significant effects on
such long timescales). At 120AU, the time taken for a particle to complete one cycle of
secular evolution (make a loop in figure \ref{fig:sec}) varies in the range 2-7$\times
10^5$ years, with larger loops taking longer. These times will also scale with particle
semi-major axis. Particles 1-12 are those with initial inclinations between 0-90$^\circ$
that are sufficient to describe the range of spatial structures as we cannot distinguish
between pro/retrograde orbits.

The particles can be split into two groups; those with low inclinations whose nodes
circulate (1--3) and those with high inclinations whose nodes librate about 90$^\circ$
(4--12). The dividing line (separatrix) between these families for the binary
eccentricity of 0.76 is $21^\circ$ when $\Omega=90^\circ$ \citep[or
$270^\circ$,][]{2010MNRAS.401.1189F}. While particles in the first group have
$i<21^\circ$ when $\Omega=90^\circ$, their inclinations when $\Omega=0^\circ$ (or
$180^\circ$) can be as high as $90^\circ$. Thus, particles near the separatrix will sweep
out an entire spherical shell during their secular evolution. Similarly, particles near
the separatrix but in the second group also sweep out a spherical shell, though the
orbital evolution is different.

\begin{figure*}
  \begin{center}
    \includegraphics[width=1\textwidth]{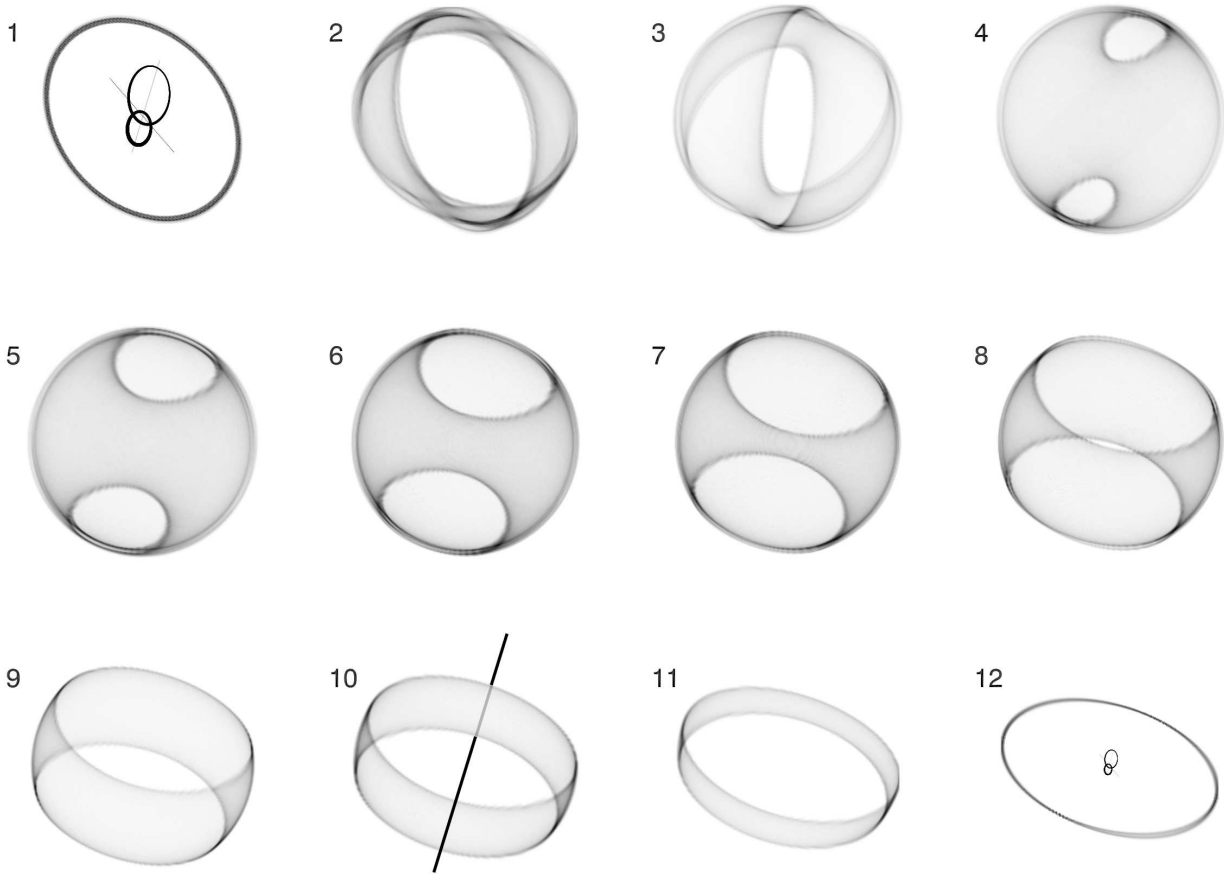}
    \caption{Debris structures derived from the secular evolution of particles 1-12 in
      Figure \ref{fig:sec} as they would be seen in the 99 Her system. In panels 1 and 12
      we have included the binary orbit from Figure \ref{fig:sys}. The orbit is not to
      scale in panel 1 for better visualisation, but is in panel 12. Panel 10 includes a
      line along the binary pericenter direction that is obscured by the ring to show the
      orientation. In panels 1-3 particles nodes circulate and the binary orbital plane
      is the plane of symmetry. In panel 1, the coplanar case, the position angle of the
      disk is aligned with the line of nodes. In panel 3 the binary is nearly surrounded
      by a broad shell of particles. In panels 4-12 particles nodes librate and the plane
      of symmetry is perpendicular to the binary pericenter direction. In panels 7-12 the
      observed position angle is perpendicular to the binary
      pericenter.}\label{fig:swift}
  \end{center}
\end{figure*}  

To visualise the structures swept out by these families of particles due to secular
perturbations, Figure \ref{fig:swift} shows the resulting debris structures for particles
that follow each of the trajectories 1-12 from Figure \ref{fig:sec} (left to right and
down). The structures are oriented as they would be seen on the sky in the 99 Her system
(i.e. have the same orientation with respect to the binary orbit shown in Figure
\ref{fig:sys}). Each structure was generated by taking the relevant particle at each time
step and spawning 1000 additional particles spread randomly around the orbit. This
process was repeated for every time step, thus building up the spatial density of a
family of particles that follow a specific curve in Figure \ref{fig:sec}. These
structures are optically thin, which makes interpreting them somewhat difficult. We have
included a scaled version of the binary orbit from Figure \ref{fig:sys} in some panels in
an attempt to make the orientations clearer.

The first (top left) panel shows a circular orbit coplanar with the binary. The PA is the
binary line of nodes, and Figure \ref{fig:copl} shows why a disk in the plane of the
binary is not a satisfactory match to the observations. The second and third panels are
still symmetric about the binary orbital plane, but have a wider range of inclinations
and are an even poorer match to the observations. Panel 3 shows that while particle
inclinations are restricted for $\Omega=90,270^\circ$, they can be large for
$\Omega=0,180^\circ$ and result in a ``butterfly'' structure when viewed down the binary
pericenter direction.

The remaining panels are for particles 4-12, whose nodes librate and for which the plane
of symmetry is perpendicular to the binary pericenter direction. In panel 4 the range of
nodes and inclinations is so large that a particle sweeps out nearly an entire spherical
shell during a cycle of secular evolution (i.e. the particle is near the
separatrix). This range decreases as the initial inclination nears a polar orbit, at
which point the orbital elements do not evolve and the resulting structure appears in
panel 12 as a simple ring. The key difference from the ring in panel 1 is that this
ring's position angle is perpendicular to the sky projection of the binary pericenter
direction, and as noted in \S\ref{s:basic} is therefore similar to the observed PA in the
PACS images.

Secular perturbations from the binary also affect the long term evolution of particle
eccentricities and pericenter longitudes. These effects are taken into account by our
$n$-body approach. However, we noticed that the eccentricities imposed (``forced'') on
the particles are lower than would be expected for a lower mass companion. Further
$n$-body simulations of coplanar particles show that for 99 Her with a mass ratio of 0.49
the forced eccentricity at 120AU is about 0.03, but if the mass ratio were 0.05 the
forced $e$ is 0.1.

This lack of significant eccentricity forcing is visible by its absence in Figure
\ref{fig:swift}, where the structures would be much broader if there were a large range
of particle eccentricities. For example, if the mass of the secondary in the 99 Her
system were significantly smaller, the model in panel 1 would become broader and offset
from the binary center of mass, resulting in a small pericenter glow effect.

This dependence suggests that a circumbinary disk's structure may help constrain the
binary mass ratio in cases where it is uncertain. However, we cannot apply this idea to
make a better estimate of the 99 Her mass ratio because the PACS observations do not have
enough resolution. In addition, at high inclinations the particle behaviour is more
complicated, because polar particles switch between pro and retrograde orbits and do not
follow simple circles in complex eccentricity space.

\subsubsection{Polar ring model}

\begin{figure*}
  \begin{center}
    \hspace{-0.5cm} \includegraphics[width=1\textwidth]{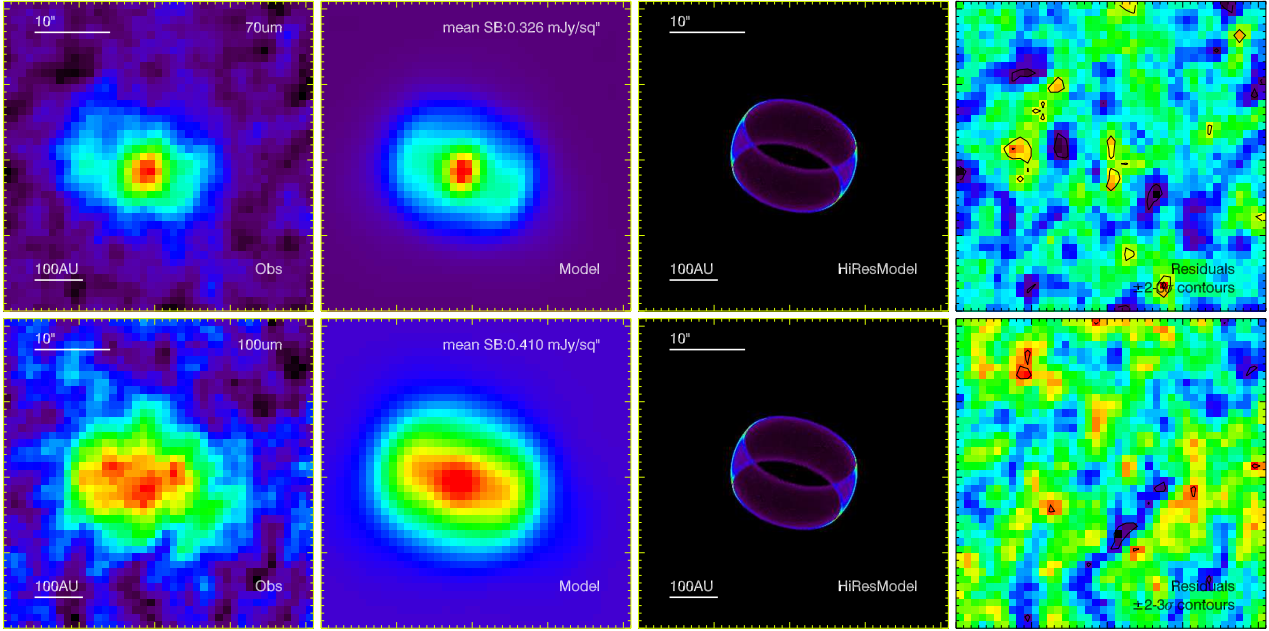}
    \caption{Polar ring model compared to the PACS 70 (top row) and 100$\mu$m (bottom
      row) observations. Layout is as for Figure \ref{fig:copl}.}\label{fig:polar}
  \end{center}
\end{figure*}

We now use the models from Figure \ref{fig:swift} to fit the PACS observations. The model
has only seven free parameters; the particle semi-major axis and initial inclination, the
surface area of dust, and the same four RA/Dec positions. The dust temperature is fixed
to 49K. Using a semi-major axis of 120AU, each panel was compared to the PACS images,
setting the surface area in grains for each model to obtain the least residual
structure. Of these we found that panel 9 was the best fit, shown in Figure
\ref{fig:polar}. These particles follow near-polar orbits so we call this model a ``polar
ring.''  We find $\chi^2=3202$. In terms of $\chi^2$ the results for panels 8 and 10 are
similar, but slightly higher. The uncertainty in the initial inclination is therefore
about 10$^\circ$, and for the semi-major axis about 10AU. This model is much better than
the coplanar model of Figure \ref{fig:copl}, with no overlapping residual structure at 70
and 100$\mu$m. The particles likely occupy a wider range of orbits than a single
semi-major axis with some non-zero eccentricity, which may account from some minor
(2$\sigma$) structure in the residuals at the disk ansae at 70$\mu$m. However, given that
this model stems directly from the secular evolution, has very few free parameters, and
accounts for the structure in all PACS images, we consider it a plausible explanation.

\subsection{Transient ring model}\label{s:ring}

A simple circular ring is a natural model to fit to the observations. This model has
eight free parameters, with the width of the ring fixed at 20AU and the opening angle
fixed to 5$^\circ$. As expected from the simple analysis in \S\ref{s:basic} the position
angle of this ring is not aligned with the binary line of nodes, and is therefore
misaligned with the binary orbit.

The interpretation depends on the orientation of the best fit. A misaligned ring with
polar orbits and the correct line of nodes would be considered further evidence in favour
of the above polar ring model. A ring with a non-polar orientation will be spread into a
broad structure like one of the panels in Figure \ref{fig:swift} by secular
perturbations. The ring cannot be long-lived and could therefore be the aftermath of a
recent collision, seen after the collision products have sheared into a ring, but before
secular perturbations destroy the ring structure. Thus we call this model a ``transient
ring.''

\begin{figure*}
  \begin{center}
    \hspace{-0.5cm} \includegraphics[width=1\textwidth]{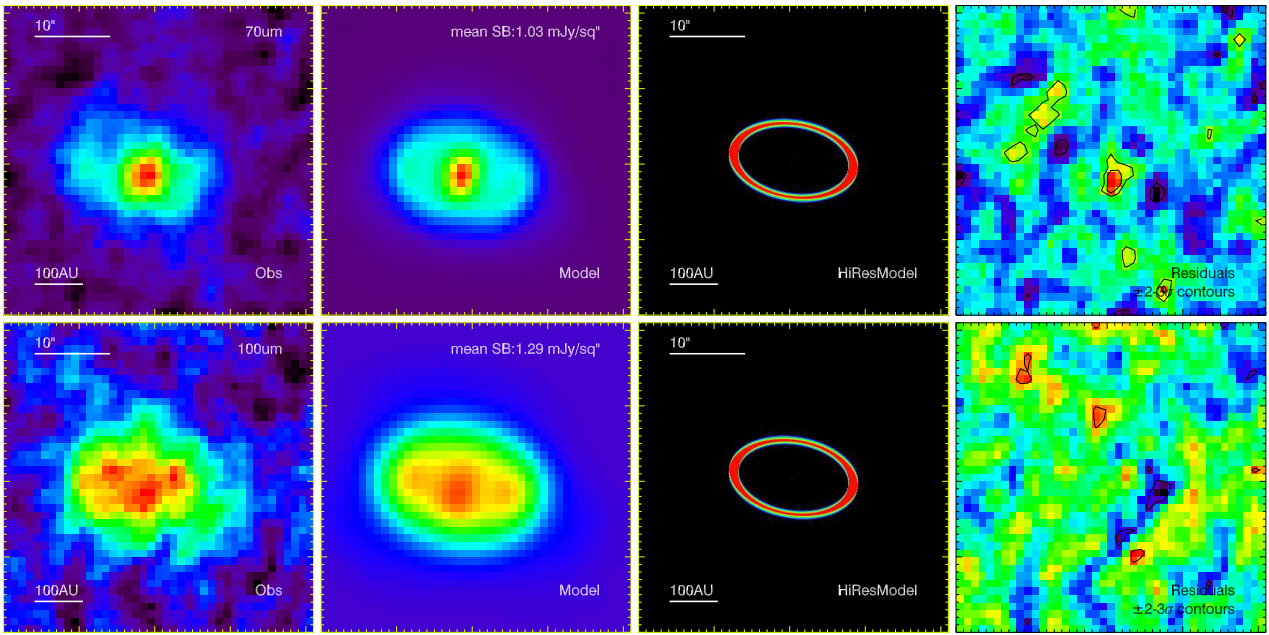}
    \caption{Transient ring model compared to the PACS 70 (top row) and 100$\mu$m (bottom
      row) observations. Layout is as for Figure \ref{fig:copl}.}\label{fig:ring}
  \end{center}
\end{figure*}  

This model is shown in Figure \ref{fig:ring}, and is a reasonable match to the PACS
observations. However, the residuals at 70$\mu$m show that the ring produces a structure
that is slightly too elliptical, compared to the more rectangular structure that is
observed and reproduced by the polar ring. This model also has less emission at the
stellar position than is observed. For this model $\chi^2= 3304$. The disk is inclined
53$^\circ$ from face-on and the PA is 81$^\circ$. The uncertainties are similar to those
derived for the Gaussian fits in \S\ref{s:basic}. The minimum relative inclination
between the disk and binary orbital planes is therefore 32 degrees, with a line of nodes
with respect to the binary orbit of 139$^\circ$.

However, the inclination between the disk and binary plane could also be 87$^\circ$ if
the disk were mirrored in the sky plane, which means that the particles have near-polar
orbits. These orbits are nearly the same as panel 12 of Figure \ref{fig:swift} (the
narrow polar ring) because the line of nodes with respect to the binary orbit is
276$^\circ$.

These two interpretations correspond to two points in Figure \ref{fig:sec}, shown as
crosses. Over time the particles would spread around to make two more lines similar to
those drawn. The particles in the lower inclination case are close to the separatrix, and
would therefore sweep out a near-spherical shell like panel 4 of Figure
\ref{fig:swift}. In this case, the long term evolution produces structures that have the
wrong position angle and are a poor match to the observations. The higher inclination
case is very nearly a polar ring and would look very similar to panel 11. Such a result
is expected because we found above that the polar ring model works well, and argues in
favour of the polar-ring interpretation.

We can in fact improve this simple ring model by increasing the total disk opening angle
(i.e. allowing a larger range of inclinations), which emulates the range of inclinations
that result from the secular evolution. We find a best fit when the particle inclinations
are 25$^\circ$ (total opening angle of 50$^\circ$), where $\chi^2=3210$. This model looks
very similar to the preferred polar ring model above, but is not generated in the same
way, and will therefore change somewhat due to secular perturbations over time because
the disk is not perfectly polar.


\section{Discussion}\label{sec:disc}

We strongly favour the polar ring model as the best explanation of the disk structure
surrounding 99 Her. The polar ring is stable for the stellar lifetime, and takes the
secular dynamics into account. The transient ring model, where the disk orientation is
not fixed, also finds that the disk particles can have polar orbits. However, because the
ring could be mirrored in the sky plane and appear the same, the ring could be misaligned
with the binary orbital plane by about 30$^\circ$. Based on $\chi^2$ and the residuals
the polar ring is marginally preferable over the transient ring. However, given that a
more realistic model with a range of particle radii and inclinations could improve the
fit in each case, we do not assign much importance to the relatively small $\chi^2$
differences. Instead, we consider several constraints on the collisional evolution that
argue against the transient ring interpretation.

By considering the timescales for collisions and secular evolution, we can estimate the
likelihood of observing the products of a collision as a transient ring before it is
spread into a broader structure. Based on the observed radius and total cross-sectional
area, the collisional lifetime of grains just above the blowout size is about a million
years \citep{2007ApJ...658..569W}. The emission could last much longer if larger objects
exist in the size distribution, and the lifetime scales with the maximum size as
$\sqrt{D_{\rm max}/D_{\rm bl}}$, so depends on the size of the largest fragment created
in the collision. If the largest fragments are at least 1mm in size the lifetime is at
least 50Myr, and we would expect the collisional cascade to be detectable for this length
of time. The secular precession timescale is about 0.5Myr, and it is reasonable to assume
that the ring structure would be erased by secular perturbations within 10 secular
timescales. Thus, the collisional products would be observable as a ring for only
5Myr. Because the collision time is longer than the secular time, the collision products
would spend at most a tenth of their lifetime appearing as a misaligned ring, and the
remainder as a broader structure. That is, assuming such a collision did occur, we have
less than a 1:10 chance of observing the collision products as a ring that looks like the
\emph{Herschel} observations.\footnote{Had we found that an eccentric ring could explain
  the data, the same argument applied to ring spreading by pericenter precession would
  apply, with the same 1:10 result. The particles' pericenter directions are unlikely to
  be maintained through forcing by a third (circumbinary) body as for a standard
  pericenter glow model, because the perturbing body would be subject to the same
  pericenter precession.}

While 1:10 is not unreasonable, this estimate does not consider the object that must be
destroyed to generate the observed dust or the plausibility of a 1mm maximum size. To
produce the observed fractional luminosity, a parent body of at least 600km in diameter
must be broken into blowout sized grains. With the more realistic assumption that the
collision produced a range of grain sizes, the parent body must be larger, about 2000km
if grains were as large as the 1mm assumed above (assuming $q=11/6$). Under the more
realistic assumption of a wide range of fragment sizes, up to 1km say, the parent body
would need to be roughly Earth-sized. However, for such large fragments the dust lifetime
would be 50Gyr and the chance of observing the structure as a ring very unlikely
(1:10,000).

We can estimate the ability of collisions to smash large objects into small pieces by
considering their strength and possible collision velocities. The specific energy needed
for catastrophic disruption, where the largest collision product is half the original
mass (i.e. very large), is roughly $10^{11}$erg/g for objects 2000km in size
\citep{2009ApJ...691L.133S}. The energy needed to disrupt an object so that the collision
products are all very small must be larger. The maximum collision energy possible for
circular orbits is for a collision between two equal sized objects on pro and retrograde
orbits. The collision energy assuming such an impact at twice the orbital velocity of a
few km/s at 100AU is a few $10^{10}$erg/g. Therefore, only in the most optimistic
(highest velocity) case is the collision energy sufficient to catastrophically disrupt
2000km objects. In the even of a disruption, the lifetime of the collision products will
be very long because the largest remnant is about 1000km in size. In the more realistic
case where collision velocities are set by object eccentricities and inclinations,
disruption of large objects at large semi-major axes is even more difficult. This
difficulty, combined with the smaller amount of starlight intercepted at such distances
means that single collisions only produce a minor increase over the background level of
dust \citep{2005AJ....130..269K}. These probability and collision arguments suggest that
a single collision is an extremely unlikely explanation for the origin of the observed
dust.

The polar ring model does not have these issues. The secular evolution of particles in
the 99 Her system means that particles on polar orbits suffer only minor changes in
inclination and node (Fig \ref{fig:sec}). These orbits are therefore stable over the
stellar lifetime so the dust could be the steady state collision products of the polar
planetesimal belt. Initial misalignment is therefore the only special requirement for the
polar ring model. The excellent agreement between the PACS data and a simple model
generated by particles on these orbits argues strongly in favour of this interpretation.

The question is then shifted to one of the origin of the misalignment. Most binaries are
thought to form through fragmentation and subsequent accretion during collapse of a
molecular cloud \citep[for a recent review see][]{2007prpl.conf..133G}. The resulting
binary systems should be aligned with their protoplanetary disks when the separations are
of order tens of AU \citep{2000MNRAS.317..773B}. Given the 16AU separation of the 99 Her
system, it therefore seems that interactions during the subsequent phase of dynamical
cluster evolution are a more likely cause of a misaligned disk.

There are several ways that such a configuration could arise from interactions in a young
stellar cluster. A close encounter in which a binary captures some material from the
outer regions of a circumstellar disk hosted by another star seems possible. This
``disk exchange'' scenario requires an encounter where the binary pericenter is
perpendicular to the circumstellar disk plane, and that the encounter distance and
geometry captures material into orbits similar to those observed for the debris disk
(e.g. most likely a prograde rather than retrograde encounter).

An alternative scenario is a stellar exchange reaction, where a binary encounters a
single star that harbours a circumstellar disk. During the exchange one of the binary
components is captured by the originally single star, and the other leaves
\citep[e.g.][]{2011arXiv1109.2007M}. The post-encounter configuration is then a binary
surrounded by a circumbinary disk. If the binary pericenter direction were perpendicular
to the disk plane it could represent a young analogue of the 99 Her system. Such an
encounter would require that the disk is not irreparably damaged by large stellar
excursions during the exchange \citep{2011arXiv1109.2007M}, but may also present a way to
clear inner disk regions, thus providing a possible reason that the 99 Her disk resides
at 120AU and not closer, where it could still be stable (see \S~\ref{s:dyn}).

Both scenarios require some degree of tuning; the encounters must happen with specific
geometries to produce the observed relative binary and disk orientations. However,
differences in the surface brightness between the different models in Figure
\ref{fig:swift} mean there could be some selection bias towards more disk-like
structures. The advantage of the disk exchange scenario is that the cross section for
interaction at a distance of about 100AU is much higher than for stellar exchange, which
would need to have an encounter distance similar to the binary semi-major axis. With a
factor of about ten difference in the encounter impact parameter for each scenario, the
close encounter is therefore about 100 times more likely than the exchange (ignoring
other constraints on geometry, configuration etc.).

In the absence of detailed simulations of encounter outcomes, some data exist to help
distinguish between these two scenarios. The minimum inclination of the stellar pole for
the 99 Her primary relative to the binary orbital plane is $20 \pm 10^\circ$
\citep{1994AJ....107..306H}. The inclination difference is therefore different from the
binary plane with 95\% confidence, and is a hint that the system may be the result of an
exchange. However, the scatter in inclination differences for binaries with separations
similar to that of 99 Her is about 20$^\circ$ \citep{1994AJ....107..306H}, which may
indicate that systems with this separation are in fact aligned and the uncertainties were
underestimated, or that this scatter is the intrinsic level of misalignment at these
separations.

Though 99 Herculis is the first clear case of misalignment between binary and disk
planes, the GG Tauri system may show a similar signature. The GG Tau system consists of
an Aa/Ab binary surrounded by a circumbinary ring
\citep{1999A&A...348..570G,2011A&A...528A..81P}, and a more distant Ba/Bb pair that may
be bound \citep{2006A&A...446..137B}. It is not clear if the inner binary is misaligned
with the circumbinary disk, but is suggested because if they are aligned the ring's inner
edge is too distant to be set by the binary
\citep{2005A&A...439..585B,2006A&A...446..137B,2011A&A...530A.126K}. However, there could
also be problems if they are misaligned, because the expected disk scale height due to
perturbations from the binary may be inconsistent with observations
\citep{2006A&A...446..137B}. Though uncertain, the possible misalignment between the
binary and ring planes shows that GG Tau could be a young analogue of 99 Her-like
systems.

\section{Summary}

We have modelled the resolved circumbinary debris disk in the 99 Her system. This disk is
unusual because it appears misaligned with the binary plane. It can be explained as
either an inclined transient ring due to a recent collision, or more likely a ring of
polar orbits. The transient ring is shown to be implausible from collisional
arguments. While the inclined ring cannot exist on long (secular) timescales, the polar
ring can.

There appear to be two possible formation scenarios for the polar ring model, which both
invoke stellar encounters. The binary may have captured material from another stars'
circumstellar disk, or a new binary may have formed in a stellar exchange where one of
the systems already contained a circumstellar disk.

While many binary and multiple systems are known to have debris disks, none are resolved
and have orbits characterised as well as 99 Herculis. Future efforts should characterise
this system further to test our interpretation and attempt to find more examples. A
sample of resolved circumbinary disks would test whether disk-binary misalignment is a
common outcome of star formation and cluster evolution, with implications for planetary
systems around both single and binary stars.

\section*{Acknowledgments}

We are grateful to the referee for a thorough reading of the manuscript, especially for
noting that previous 99 Her visual orbits have the wrong ascending node. This research
has made use of the Washington Double Star Catalog maintained at the U.S. Naval
Observatory, and the SwiftVis $n$-body visualisation software developed by Mark Lewis. We
also thank Herve Beust for use of the HJS code, and Paul Harvey for comments on a draft
of this article.


\begin{thebibliography}{59}
\expandafter\ifx\csname natexlab\endcsname\relax\def\natexlab#1{#1}\fi
\expandafter\ifx\csname href\endcsname\relax
  \def\href#1#2{}\fi
\expandafter\ifx\csname urllinklabel\endcsname\relax
  \def\urllinklabel{[LINK]}\fi
\expandafter\ifx\csname adsurllinklabel\endcsname\relax
  \def\adsurllinklabel{[ADS]}\fi

\bibitem[{{Abt} \& {Willmarth}(2006)}]{2006ApJS..162..207A}
{Abt}, H.~A. \& {Willmarth}, D. 2006, \apjs, 162, 207
 \href{http://adsabs.harvard.edu/abs/2006ApJS..162..207A}{\adsurllinklabel}

\bibitem[{{Adelman} {et~al.}(2000){Adelman}, {Caliskan}, {Kocer}, {Cay}, \&
  {Gokmen Tektunali}}]{2000MNRAS.316..514A}
{Adelman}, S.~J., {Caliskan}, H., {Kocer}, D., {Cay}, I.~H., \& {Gokmen
  Tektunali}, H. 2000, \mnras, 316, 514
 \href{http://adsabs.harvard.edu/abs/2000MNRAS.316..514A}{\adsurllinklabel}

\bibitem[{{Amblard} {et~al.}(2010)}]{2010A&A...518L...9A}
{Amblard}, A. {et~al.} 2010, \aap, 518, L9
 \href{http://adsabs.harvard.edu/abs/2010A&A...518L...9A}{\adsurllinklabel}

\bibitem[{{Augereau} {et~al.}(2001){Augereau}, {Nelson}, {Lagrange},
  {Papaloizou}, \& {Mouillet}}]{2001A&A...370..447A}
{Augereau}, J.~C., {Nelson}, R.~P., {Lagrange}, A.~M., {Papaloizou}, J.~C.~B.,
  \& {Mouillet}, D. 2001, \aap, 370, 447
 \href{http://adsabs.harvard.edu/abs/2001A&A...370..447A}{\adsurllinklabel}

\bibitem[{{Bate} {et~al.}(2000){Bate}, {Bonnell}, {Clarke}, {Lubow}, {Ogilvie},
  {Pringle}, \& {Tout}}]{2000MNRAS.317..773B}
{Bate}, M.~R., {Bonnell}, I.~A., {Clarke}, C.~J., {Lubow}, S.~H., {Ogilvie},
  G.~I., {Pringle}, J.~E., \& {Tout}, C.~A. 2000, \mnras, 317, 773
 \href{http://adsabs.harvard.edu/abs/2000MNRAS.317..773B}{\adsurllinklabel}

\bibitem[{{Beust}(2003)}]{2003A&A...400.1129B}
{Beust}, H. 2003, \aap, 400, 1129
 \href{http://adsabs.harvard.edu/abs/2003A&A...400.1129B}{\adsurllinklabel}

\bibitem[{{Beust} \& {Dutrey}(2005)}]{2005A&A...439..585B}
{Beust}, H. \& {Dutrey}, A. 2005, \aap, 439, 585
 \href{http://adsabs.harvard.edu/abs/2005A&A...439..585B}{\adsurllinklabel}

\bibitem[{{Beust} \& {Dutrey}(2006)}]{2006A&A...446..137B}
---. 2006, \aap, 446, 137
 \href{http://adsabs.harvard.edu/abs/2006A&A...446..137B}{\adsurllinklabel}

\bibitem[{{Boesgaard} {et~al.}(2004){Boesgaard}, {McGrath}, {Lambert}, \&
  {Cunha}}]{2004ApJ...606..306B}
{Boesgaard}, A.~M., {McGrath}, E.~J., {Lambert}, D.~L., \& {Cunha}, K. 2004,
  \apj, 606, 306
 \href{http://adsabs.harvard.edu/abs/2004ApJ...606..306B}{\adsurllinklabel}

\bibitem[{{Churcher} {et~al.}(2011){Churcher}, {Wyatt}, {Duch{\^e}ne},
  {Sibthorpe}, {Kennedy}, {Matthews}, {Kalas}, {Greaves}, {Su}, \&
  {Rieke}}]{2011MNRAS.417.1715C}
{Churcher}, L.~J., {Wyatt}, M.~C., {Duch{\^e}ne}, G., {Sibthorpe}, B.,
  {Kennedy}, G., {Matthews}, B.~C., {Kalas}, P., {Greaves}, J., {Su}, K., \&
  {Rieke}, G. 2011, \mnras, 417, 1715
 \href{http://adsabs.harvard.edu/abs/2011MNRAS.417.1715C}{\adsurllinklabel}

\bibitem[{{Dommanget} \& {Nys}(2002)}]{2002yCat.1274....0D}
{Dommanget}, J. \& {Nys}, O. 2002, VizieR Online Data Catalog, 1274, 0
 \href{http://adsabs.harvard.edu/abs/2002yCat.1274....0D}{\adsurllinklabel}

\bibitem[{{Doolin} \& {Blundell}(2011)}]{2011arXiv1108.4144D}
{Doolin}, S. \& {Blundell}, K.~M. 2011, ArXiv e-prints, (1108.4144)
 \href{http://adsabs.harvard.edu/abs/2011arXiv1108.4144D}{\adsurllinklabel}

\bibitem[{{Farago} \& {Laskar}(2010)}]{2010MNRAS.401.1189F}
{Farago}, F. \& {Laskar}, J. 2010, \mnras, 401, 1189
 \href{http://adsabs.harvard.edu/abs/2010MNRAS.401.1189F}{\adsurllinklabel}

\bibitem[{{Fruchter} \& {Hook}(2002)}]{2002PASP..114..144F}
{Fruchter}, A.~S. \& {Hook}, R.~N. 2002, \pasp, 114, 144
 \href{http://adsabs.harvard.edu/abs/2002PASP..114..144F}{\adsurllinklabel}

\bibitem[{{Goodwin} {et~al.}(2007){Goodwin}, {Kroupa}, {Goodman}, \&
  {Burkert}}]{2007prpl.conf..133G}
{Goodwin}, S.~P., {Kroupa}, P., {Goodman}, A., \& {Burkert}, A. 2007,
  Protostars and Planets V, 133
 \href{http://adsabs.harvard.edu/abs/2007prpl.conf..133G}{\adsurllinklabel}

\bibitem[{{Gratton} {et~al.}(1996){Gratton}, {Carretta}, \&
  {Castelli}}]{1996yCat..33140191G}
{Gratton}, R.~G., {Carretta}, E., \& {Castelli}, F. 1996, VizieR Online Data
  Catalog, 331, 40191
 \href{http://adsabs.harvard.edu/abs/1996yCat..33140191G}{\adsurllinklabel}

\bibitem[{{Griffin} {et~al.}(2010)}]{2010A&A...518L...3G}
{Griffin}, M.~J. {et~al.} 2010, \aap, 518, L3
 \href{http://adsabs.harvard.edu/abs/2010A&A...518L...3G}{\adsurllinklabel}

\bibitem[{{Guilloteau} {et~al.}(1999){Guilloteau}, {Dutrey}, \&
  {Simon}}]{1999A&A...348..570G}
{Guilloteau}, S., {Dutrey}, A., \& {Simon}, M. 1999, \aap, 348, 570
 \href{http://adsabs.harvard.edu/abs/1999A&A...348..570G}{\adsurllinklabel}

\bibitem[{{Hale}(1994)}]{1994AJ....107..306H}
{Hale}, A. 1994, \aj, 107, 306
 \href{http://adsabs.harvard.edu/abs/1994AJ....107..306H}{\adsurllinklabel}

\bibitem[{{Hartkopf} {et~al.}(2001){Hartkopf}, {Mason}, \&
  {Worley}}]{2001AJ....122.3472H}
{Hartkopf}, W.~I., {Mason}, B.~D., \& {Worley}, C.~E. 2001, \aj, 122, 3472
 \href{http://adsabs.harvard.edu/abs/2001AJ....122.3472H}{\adsurllinklabel}

\bibitem[{{Hauck} \& {Mermilliod}(1997)}]{1997yCat.2215....0H}
{Hauck}, B. \& {Mermilliod}, M. 1997, VizieR Online Data Catalog, 2215, 0
 \href{http://adsabs.harvard.edu/abs/1997yCat.2215....0H}{\adsurllinklabel}

\bibitem[{{Henry} \& {McCarthy}(1993)}]{1993AJ....106..773H}
{Henry}, T.~J. \& {McCarthy}, Jr., D.~W. 1993, \aj, 106, 773
 \href{http://adsabs.harvard.edu/abs/1993AJ....106..773H}{\adsurllinklabel}

\bibitem[{{Ishihara} {et~al.}(2010)}]{2010A&A...514A...1I}
{Ishihara}, D. {et~al.} 2010, \aap, 514, A1
 \href{http://adsabs.harvard.edu/abs/2010A&A...514A...1I}{\adsurllinklabel}

\bibitem[{{Kenyon} \& {Bromley}(2005)}]{2005AJ....130..269K}
{Kenyon}, S.~J. \& {Bromley}, B.~C. 2005, \aj, 130, 269
 \href{http://adsabs.harvard.edu/abs/2005AJ....130..269K}{\adsurllinklabel}

\bibitem[{{Koerner} {et~al.}(2010){Koerner}, {Kim}, {Trilling}, {Larson},
  {Cotera}, {Stapelfeldt}, {Wahhaj}, {Fajardo-Acosta}, {Padgett}, \&
  {Backman}}]{2010ApJ...710L..26K}
{Koerner}, D.~W., {Kim}, S., {Trilling}, D.~E., {Larson}, H., {Cotera}, A.,
  {Stapelfeldt}, K.~R., {Wahhaj}, Z., {Fajardo-Acosta}, S., {Padgett}, D., \&
  {Backman}, D. 2010, \apjl, 710, L26
 \href{http://adsabs.harvard.edu/abs/2010ApJ...710L..26K}{\adsurllinklabel}

\bibitem[{{K{\"o}hler}(2011)}]{2011A&A...530A.126K}
{K{\"o}hler}, R. 2011, \aap, 530, A126
 \href{http://adsabs.harvard.edu/abs/2011A&A...530A.126K}{\adsurllinklabel}

\bibitem[{{Kozai}(1962)}]{1962AJ.....67..591K}
{Kozai}, Y. 1962, \aj, 67, 591
 \href{http://adsabs.harvard.edu/abs/1962AJ.....67..591K}{\adsurllinklabel}

\bibitem[{{Li} \& {Greenberg}(1997)}]{1997A&A...323..566L}
{Li}, A. \& {Greenberg}, J.~M. 1997, \aap, 323, 566
 \href{http://adsabs.harvard.edu/abs/1997A&A...323..566L}{\adsurllinklabel}

\bibitem[{{Lidov}(1962)}]{1962P&SS....9..719L}
{Lidov}, M.~L. 1962, \planss, 9, 719
 \href{http://adsabs.harvard.edu/abs/1962P&SS....9..719L}{\adsurllinklabel}

\bibitem[{{Mason} {et~al.}(2011){Mason}, {Wycoff}, {Hartkopf}, {Douglass}, \&
  {Worley}}]{2011yCat....102026M}
{Mason}, B.~D., {Wycoff}, G.~L., {Hartkopf}, W.~I., {Douglass}, G.~G., \&
  {Worley}, C.~E. 2011, VizieR Online Data Catalog, 1, 2026
 \href{http://adsabs.harvard.edu/abs/2011yCat....102026M}{\adsurllinklabel}

\bibitem[{{Matthews} {et~al.}(2010)}]{2010A&A...518L.135M}
{Matthews}, B.~C. {et~al.} 2010, \aap, 518, L135
 \href{http://adsabs.harvard.edu/abs/2010A&A...518L.135M}{\adsurllinklabel}

\bibitem[{{Mermilliod}(2006)}]{2006yCat.2168....0M}
{Mermilliod}, J.~C. 2006, VizieR Online Data Catalog, 2168, 0
 \href{http://adsabs.harvard.edu/abs/2006yCat.2168....0M}{\adsurllinklabel}

\bibitem[{{Moeckel} \& {Goddi}(2011)}]{2011arXiv1109.2007M}
{Moeckel}, N. \& {Goddi}, C. 2011, ArXiv e-prints, (1109.2007)
 \href{http://adsabs.harvard.edu/abs/2011arXiv1109.2007M}{\adsurllinklabel}

\bibitem[{{Moshir} \& {et al.}(1990)}]{1990IRASF.C......0M}
{Moshir}, M. \& {et al.} 1990, in IRAS Faint Source Catalogue, version 2.0
  (1990), 0
 \href{http://adsabs.harvard.edu/abs/1990IRASF.C......0M}{\adsurllinklabel}

\bibitem[{{Nordstr{\"o}m} {et~al.}(2004){Nordstr{\"o}m}, {Mayor}, {Andersen},
  {Holmberg}, {Pont}, {J{\o}rgensen}, {Olsen}, {Udry}, \&
  {Mowlavi}}]{2004A&A...418..989N}
{Nordstr{\"o}m}, B., {Mayor}, M., {Andersen}, J., {Holmberg}, J., {Pont}, F.,
  {J{\o}rgensen}, B.~R., {Olsen}, E.~H., {Udry}, S., \& {Mowlavi}, N. 2004,
  \aap, 418, 989
 \href{http://adsabs.harvard.edu/abs/2004A&A...418..989N}{\adsurllinklabel}

\bibitem[{{O'Brien} \& {Greenberg}(2003)}]{2003Icar..164..334O}
{O'Brien}, D.~P. \& {Greenberg}, R. 2003, \icarus, 164, 334
 \href{http://adsabs.harvard.edu/abs/2003Icar..164..334O}{\adsurllinklabel}

\bibitem[{{Ott}(2010)}]{2010ASPC..434..139O}
{Ott}, S. 2010, in Astronomical Society of the Pacific Conference Series, Vol.
  434, Astronomical Data Analysis Software and Systems XIX, ed. {Y.~Mizumoto,
  K.-I.~Morita, \& M.~Ohishi}, 139
 \href{http://adsabs.harvard.edu/abs/2010ASPC..434..139O}{\adsurllinklabel}

\bibitem[{{Perryman} \& {ESA}(1997)}]{1997ESASP1200.....P}
{Perryman}, M.~A.~C. \& {ESA}, eds. 1997, ESA Special Publication, Vol. 1200,
  {The HIPPARCOS and TYCHO catalogues. Astrometric and photometric star
  catalogues derived from the ESA HIPPARCOS Space Astrometry Mission}
 \href{http://adsabs.harvard.edu/abs/1997ESASP1200.....P}{\adsurllinklabel}

\bibitem[{{Phillips} {et~al.}(2010){Phillips}, {Greaves}, {Dent}, {Matthews},
  {Holland}, {Wyatt}, \& {Sibthorpe}}]{2010MNRAS.403.1089P}
{Phillips}, N.~M., {Greaves}, J.~S., {Dent}, W.~R.~F., {Matthews}, B.~C.,
  {Holland}, W.~S., {Wyatt}, M.~C., \& {Sibthorpe}, B. 2010, \mnras, 403, 1089
 \href{http://adsabs.harvard.edu/abs/2010MNRAS.403.1089P}{\adsurllinklabel}

\bibitem[{{Pi{\'e}tu} {et~al.}(2011){Pi{\'e}tu}, {Gueth}, {Hily-Blant},
  {Schuster}, \& {Pety}}]{2011A&A...528A..81P}
{Pi{\'e}tu}, V., {Gueth}, F., {Hily-Blant}, P., {Schuster}, K.-F., \& {Pety},
  J. 2011, \aap, 528, A81
 \href{http://adsabs.harvard.edu/abs/2011A&A...528A..81P}{\adsurllinklabel}

\bibitem[{{Pilbratt} {et~al.}(2010){Pilbratt}, {Riedinger}, {Passvogel},
  {Crone}, {Doyle}, {Gageur}, {Heras}, {Jewell}, {Metcalfe}, {Ott}, \&
  {Schmidt}}]{2010A&A...518L...1P}
{Pilbratt}, G.~L., {Riedinger}, J.~R., {Passvogel}, T., {Crone}, G., {Doyle},
  D., {Gageur}, U., {Heras}, A.~M., {Jewell}, C., {Metcalfe}, L., {Ott}, S., \&
  {Schmidt}, M. 2010, \aap, 518, L1
 \href{http://adsabs.harvard.edu/abs/2010A&A...518L...1P}{\adsurllinklabel}

\bibitem[{{Poglitsch} {et~al.}(2010)}]{2010A&A...518L...2P}
{Poglitsch}, A. {et~al.} 2010, \aap, 518, L2
 \href{http://adsabs.harvard.edu/abs/2010A&A...518L...2P}{\adsurllinklabel}

\bibitem[{{Rieke} {et~al.}(2004)}]{2004ApJS..154...25R}
{Rieke}, G.~H. {et~al.} 2004, \apjs, 154, 25
 \href{http://adsabs.harvard.edu/abs/2004ApJS..154...25R}{\adsurllinklabel}

\bibitem[{{Rieke} {et~al.}(2008)}]{2008AJ....135.2245R}
---. 2008, \aj, 135, 2245
 \href{http://adsabs.harvard.edu/abs/2008AJ....135.2245R}{\adsurllinklabel}

\bibitem[{{Scardia} {et~al.}(2010){Scardia}, {Prieur}, {Pansecchi}, {Argyle},
  \& {Sala}}]{2010AN....331..286S}
{Scardia}, M., {Prieur}, J.-L., {Pansecchi}, L., {Argyle}, R.~W., \& {Sala}, M.
  2010, Astronomische Nachrichten, 331, 286
 \href{http://adsabs.harvard.edu/abs/2010AN....331..286S}{\adsurllinklabel}

\bibitem[{{Scardia} {et~al.}(2008){Scardia}, {Prieur}, {Pansecchi}, {Argyle},
  {Sala}, {Basso}, {Ghigo}, {Koechlin}, \& {Aristidi}}]{2008AN....329...54S}
{Scardia}, M., {Prieur}, J.-L., {Pansecchi}, L., {Argyle}, R.~W., {Sala}, M.,
  {Basso}, S., {Ghigo}, M., {Koechlin}, L., \& {Aristidi}, E. 2008,
  Astronomische Nachrichten, 329, 54
 \href{http://adsabs.harvard.edu/abs/2008AN....329...54S}{\adsurllinklabel}

\bibitem[{{S{\"o}derhjelm}(1999)}]{1999A&A...341..121S}
{S{\"o}derhjelm}, S. 1999, \aap, 341, 121
 \href{http://adsabs.harvard.edu/abs/1999A&A...341..121S}{\adsurllinklabel}

\bibitem[{{Stewart} \& {Leinhardt}(2009)}]{2009ApJ...691L.133S}
{Stewart}, S.~T. \& {Leinhardt}, Z.~M. 2009, \apjl, 691, L133
 \href{http://adsabs.harvard.edu/abs/2009ApJ...691L.133S}{\adsurllinklabel}

\bibitem[{{Takeda}(2007)}]{2007PASJ...59..335T}
{Takeda}, Y. 2007, \pasj, 59, 335
 \href{http://adsabs.harvard.edu/abs/2007PASJ...59..335T}{\adsurllinklabel}

\bibitem[{{Trilling} {et~al.}(2007){Trilling}, {Stansberry}, {Stapelfeldt},
  {Rieke}, {Su}, {Gray}, {Corbally}, {Bryden}, {Chen}, {Boden}, \&
  {Beichman}}]{2007ApJ...658.1289T}
{Trilling}, D.~E., {Stansberry}, J.~A., {Stapelfeldt}, K.~R., {Rieke}, G.~H.,
  {Su}, K.~Y.~L., {Gray}, R.~O., {Corbally}, C.~J., {Bryden}, G., {Chen},
  C.~H., {Boden}, A., \& {Beichman}, C.~A. 2007, \apj, 658, 1289
 \href{http://adsabs.harvard.edu/abs/2007ApJ...658.1289T}{\adsurllinklabel}

\bibitem[{{van Leeuwen}(2007)}]{2007ASSL..350.....V}
{van Leeuwen}, F., ed. 2007, Astrophysics and Space Science Library, Vol. 350,
  {Hipparcos, the New Reduction of the Raw Data}
 \href{http://adsabs.harvard.edu/abs/2007ASSL..350.....V}{\adsurllinklabel}

\bibitem[{{van Leeuwen}(2008)}]{2008yCat.1311....0F}
{van Leeuwen}, F. 2008, VizieR Online Data Catalog, 1311, 0
 \href{http://adsabs.harvard.edu/abs/2008yCat.1311....0F}{\adsurllinklabel}

\bibitem[{{Verrier} \& {Evans}(2008)}]{2008MNRAS.390.1377V}
{Verrier}, P.~E. \& {Evans}, N.~W. 2008, \mnras, 390, 1377
 \href{http://adsabs.harvard.edu/abs/2008MNRAS.390.1377V}{\adsurllinklabel}

\bibitem[{{Verrier} \& {Evans}(2009)}]{2009MNRAS.394.1721V}
---. 2009, \mnras, 394, 1721
 \href{http://adsabs.harvard.edu/abs/2009MNRAS.394.1721V}{\adsurllinklabel}

\bibitem[{{Wiegert} \& {Holman}(1997)}]{1997AJ....113.1445W}
{Wiegert}, P.~A. \& {Holman}, M.~J. 1997, \aj, 113, 1445
 \href{http://adsabs.harvard.edu/abs/1997AJ....113.1445W}{\adsurllinklabel}

\bibitem[{{Wyatt} \& {Dent}(2002)}]{2002MNRAS.334..589W}
{Wyatt}, M.~C. \& {Dent}, W.~R.~F. 2002, \mnras, 334, 589
 \href{http://adsabs.harvard.edu/abs/2002MNRAS.334..589W}{\adsurllinklabel}

\bibitem[{{Wyatt} {et~al.}(1999){Wyatt}, {Dermott}, {Telesco}, {Fisher},
  {Grogan}, {Holmes}, \& {Pi{\~n}a}}]{1999ApJ...527..918W}
{Wyatt}, M.~C., {Dermott}, S.~F., {Telesco}, C.~M., {Fisher}, R.~S., {Grogan},
  K., {Holmes}, E.~K., \& {Pi{\~n}a}, R.~K. 1999, \apj, 527, 918
 \href{http://adsabs.harvard.edu/abs/1999ApJ...527..918W}{\adsurllinklabel}

\bibitem[{{Wyatt} {et~al.}(2007{\natexlab{a}}){Wyatt}, {Smith}, {Greaves},
  {Beichman}, {Bryden}, \& {Lisse}}]{2007ApJ...658..569W}
{Wyatt}, M.~C., {Smith}, R., {Greaves}, J.~S., {Beichman}, C.~A., {Bryden}, G.,
  \& {Lisse}, C.~M. 2007{\natexlab{a}}, \apj, 658, 569
 \href{http://adsabs.harvard.edu/abs/2007ApJ...658..569W}{\adsurllinklabel}

\bibitem[{{Wyatt} {et~al.}(2007{\natexlab{b}}){Wyatt}, {Smith}, {Su}, {Rieke},
  {Greaves}, {Beichman}, \& {Bryden}}]{2007ApJ...663..365W}
{Wyatt}, M.~C., {Smith}, R., {Su}, K.~Y.~L., {Rieke}, G.~H., {Greaves}, J.~S.,
  {Beichman}, C.~A., \& {Bryden}, G. 2007{\natexlab{b}}, \apj, 663, 365
 \href{http://adsabs.harvard.edu/abs/2007ApJ...663..365W}{\adsurllinklabel}

\end{thebibliography}

\end{document}